\documentclass[11pt]{article}
\usepackage[latin1]{inputenc}
\usepackage{latexsym}
\usepackage{amstext}
\usepackage{amsxtra}
\usepackage{epsf}

\title{Coherence properties of a continuous atom laser}
\author{Y. Castin\thanks{E-mail: yvan.castin@lkb.ens.fr}, R. Dum and E. Mandonnet \\
Lab. Kastler Brossel de l'ENS,  24 rue Lhomond, \\
75 231 Paris Cedex 5,  France \\
A. Minguzzi and I. Carusotto \\
Scuola Normale Superiore and INFM, Piazza dei Cavalieri 7, \\
56 126 Pisa , Italy}
\date{}

\begin{document}
\maketitle
\begin{abstract}
We investigate the
coherence properties of an atomic beam evaporatively cooled in a magnetic
guide, assuming thermal equilibrium in the quantum degenerate
regime.
The gas experiences two-dimensional,
transverse Bose-Einstein condensation rather than a full three-dimensional
condensation because of the very elongated geometry of the magnetic guide.
First order and second order correlation functions of the atomic
field are used to characterize the coherence properties of the gas
along the axis of the guide.
The coherence length of the gas is found to be much larger than the thermal de Broglie wavelength
in the strongly quantum degenerate regime. Large intensity fluctuations
present in the ideal Bose gas model are found to be strongly reduced by
repulsive atomic interactions; this conclusion is obtained with a one-dimensional
classical field approximation valid when the temperature of the gas is much higher
than its chemical potential, $k_B T\gg |\mu|$.
\end{abstract}

\noindent{\bf Pacs:} 03.75.Fi, 42.50.-p

The first experimental achievements of Bose-Einstein condensates in atomic vapors
\cite{BEC1,BEC2,BEC3,BEC4} have opened promising perspectives for atom optics: condensates
constitute indeed atomic waves sources of much better coherence properties than the 
usual `thermal' sources like the standard magneto-optical trap.
These coherence properties have already been demonstrated experimentally: interferences
experiments between two condensates have been performed at MIT \cite{interf_MIT} and at JILA \cite{interf_JILA}, 
the first order correlation function of the atomic field has been measured in M\"unich \cite{Esslinger},
and suppression of density fluctuations (that is fluctuations in the intensity of the atomic field)
has been revealed by a measurement of the mean-field energy 
\cite{meanf} and of three-body losses \cite{three_body}.
By inducing a coherent leak of atoms out of trapped condensates several groups
have succeeded in creating pulsed or quasi-continuous `atom-lasers' \cite{laser_manip}.

For future applications the already realized `atom-lasers' may suffer from the handicap
of a low mean flux of atoms: the condensates were not experiencing any continuous loading 
of atoms, so that the coherent output of atoms terminated once the $\sim 10^6$ atoms of the
condensate were leaked out. As the repetition rate of the whole sequence is limited by the time required to form
a condensate by evaporative cooling (on the order of seconds) the resulting mean flux of atoms
is $< 10^6$ atoms/s. Several proposals have been made to refill the condensates with atoms
in a continuous way \cite{laser_theory} but they have to our knowledge not been realized
yet. Recently we proposed a different scheme, based on the evaporative cooling of an atomic beam 
\cite{cal}.

The goal of the present article is to predict the essential features of our `continuous
atom-laser' proposal. The paper is organized as follows.
In section \ref{evap} we summarize the calculations of \cite{cal}
performed to determine the required length of evaporative
cooling of the beam to reach the quantum degenerate regime. In section \ref{ideal} we discuss
the coherence properties of the beam once quantum degeneracy has been obtained, assuming that
the atomic interactions are negligible; we find that the beam has large intensity fluctuations
incompatible with expected coherence properties of an atom-laser. In section \ref{inter} we propose
a model of a one-dimensional interacting Bose gas and we construct a classical field approximation
to this model in the high temperature limit. We solve the classical field approximation
using the formal analogy between functional integrals and quantum propagators: we find that the interactions
between particles can dramatically reduce the intensity fluctuations of the beam. 
We conclude in section \ref{concl}.

\section{Evaporative cooling of an atomic beam}\label{evap}

\subsection{A continuous injection of atoms in a magnetic guide}\label{sub:param}
In the experimental scenario considered at the \'Ecole
normale sup\'erieure
the continuous source of atoms
is provided by standard laser cooling and trapping techniques taking place inside a cell. Atoms in the cell
are captured, cooled and trapped
in the $x-y$ plane using a two-dimensional magneto-optical trap, that is with laser beams in the 
$x-y$ plane and a dipolar magnetic field $\vec{B}\propto x\vec{e}_x-y\vec{e}_y$,
where $\vec{e}_{x,y}$ are unit vectors along $x,y$ axes. The atomic motion along
$z$ is controlled with a standard moving molasses technique:
the two counter-propagating laser beams along $z$ have different frequencies
so that the atoms are cooled around a non-zero mean velocity $\bar{v}_0$.

The already cold atomic beam emerging from this set-up is sent towards a magnetic guide of axis
$z$. The magnetic guide is produced by the superposition of a uniform magnetic field along
$z$ and a dipolar magnetic field $\propto x\vec{e}_x-y\vec{e}_y$. This provides a transverse confinement
of the atoms being in the right Zeeman sublevels, the atoms in the wrong Zeeman sublevels
being not trapped or even expelled. The magnetic guide provides a trapping potential
in the $x-y$ plane with a harmonic bottom that we write as
\begin{equation}
U(x,y) = \frac{1}{2} m\omega_\perp^2 (x^2+y^2)
\label{eq:pot}
\end{equation}
where $m$ is the atomic mass and $\omega_\perp$ the transverse oscillation frequency of the 
atoms.

The following parameters of the injected atoms are expected to be realistic for $^{87}$Rb atoms
\cite{cal}. The initial velocity
dispersion of the atoms is $\Delta v_0= 20$ cm/s corresponding to an initial temperature of 400 $\mu$K; such a high
temperature is  obtained after spatial compression of the cloud usually performed to increase the collision rate
in preparation of evaporative cooling. We take an injection velocity $\bar{v}_0=3\Delta v_0$ larger than the
velocity spread so that the incoming atoms form a beam. Assuming an injected flux of
atoms of $3\times 10^9$ atoms/s and an oscillation frequency $\omega_\perp
=2\pi\times 1$ kHz we find an initial on-axis thermal
density of $8\times 10^{11}$ atoms/cm$^3$. The initial phase
space density is $7\times 10^{-7}\ll 1$. 
The initial collision rate of atoms 
is related to the on-axis density $n_0$ and to the collisional cross section
$\sigma$ 
by $\gamma_{\rm coll}^{(0)}=(2/\sqrt{\pi})\, n_0\, \sigma \, \Delta v_0$. 
For the $s$-wave collisional cross-section of rubidium
($\sigma=7.6 \times 10^{-16}$~m$^2$) we get $\gamma_{\rm coll}^{(0)}
\simeq$ 100 s$^{-1}$ which is
much smaller than $\omega_\perp$:
\begin{equation}
\frac{\gamma_{\rm coll}^{(0)}}{\omega_\perp} = 0.02 \ll 1.
\label{eq:coll}
\end{equation}
The atoms have therefore the time to perform a full transverse oscillation in the trapping potential 
before experiencing a collision.

\subsection{Modeling of evaporative cooling}

We assume that the atoms are subject to evaporative cooling in the magnetic guide, e.g.\ by application
of a $z$-dependent radio-frequency flipping the atoms to untrapped or expelled Zeeman sublevels when
they are too far from $z$ axis.

The dynamics of evaporative cooling can be described by the classical Boltzmann equation on the phase space
atomic density $f(\vec{r},\vec{p})$ as long as the phase space density remains small.  The collision terms in
the Boltzmann equation are simplified by the assumption (justified for rubidium) that atomic
interactions take place in the $s$-wave only and have a constant (momentum independent) total cross-section $\sigma=8\pi a_{3d}^2$
($a_{3d}$ is the scattering length). The effect of evaporative cooling 
is modeled by setting $f(\vec{r},\vec{p})$ to
zero in the domain $x^2+y^2>\Lambda^2(z)$ for two-dimensional evaporation, or in the
domain $x^2 > \Lambda^2(z)$ for one-dimensional evaporation, where $\Lambda(z)$ is an adjustable
cut in position space.

We have first looked for an approximate analytical solution of the Boltzmann equation, adapting to
our geometry the truncated Gaussian ansatz put forward in \cite{Walraven}. As argued in \cite{cal} it is
then more rigorous to restrict to a one-dimensional evaporation scheme.
The ansatz then takes the following form:
\begin{equation}
f(\vec{r},\vec{p})= f_0(z) e^{-(\epsilon_x+\epsilon_y)/(k_B T(z))} e^{-(p_z-\bar{p}(z))^2/(2mk_B T(z))}
Y(\epsilon_{\rm evap}(z)-\epsilon_x)
\label{eq:ansatz}
\end{equation}
where $\epsilon_x, \epsilon_y$ are the sum of kinetic energy and harmonic trapping potential energy
along $x$ and $y$ respectively and $Y$ is the Heaviside function.
The ansatz assumes a local thermal equilibrium with temperature $T(z)$. The temperature depends on $z$ only,
not on $x$ and $y$ as, due to Eq.(\ref{eq:coll}), 
the mean free path of the particles $\sim \Delta v/\gamma_{\rm coll}$
is much larger than the spatial transverse extension of the gas $(k_B T/(m \omega_\perp^2))^{1/2}\sim \Delta v/\omega_\perp$:
transversally
the gas is in the so-called collisionless regime. For the same reason the truncation of $f$ in
position space is replaced by a truncation in energy space, with 
\begin{equation}
\epsilon_{\rm evap} (z) = \frac{1}{2} m\omega_\perp^2 \Lambda^2(z).
\end{equation}
Knowing the energy of the particle along $x$ allows to calculate the maximal excursion of
the trajectory along $x$ (as atoms have in general the time to perform a full harmonic
oscillation before experiencing a collision); if this maximal excursion exceeds $\Lambda(z)$
the particle is evaporated.

There are {\it a priori} three unknown functions of $z$ in the ansatz Eq.(\ref{eq:ansatz}): (i) the normalization
factor $f_0(z)$ or equivalently the linear density $\rho_{\rm lin}(z)$, (ii) the mean momentum $\bar{p}(z)$ of the gas along $z$,
and (iii) the temperature of the gas $T(z)$.

By multiplying 
Boltzmann's equation by (i) unity, (ii) the momentum $p_z$, and (iii) the kinetic energy along $z$,
$p_z^2/(2m)$, and by integrating over $x,y,p_x,p_y,p_z$ one gets three hydrodynamic type
equations for $\rho_{\rm lin}(z)$, $\bar{p}(z)$ and $k_B T(z)$. 
These three equations contain the usual 
equations expressing  conservation
of probability, of momentum and of energy, plus
extra terms describing the loss of particles, the change of momentum and energy
due to the evaporation. 

\begin{figure}[htb]
\epsfxsize=12cm \centerline{\epsfbox{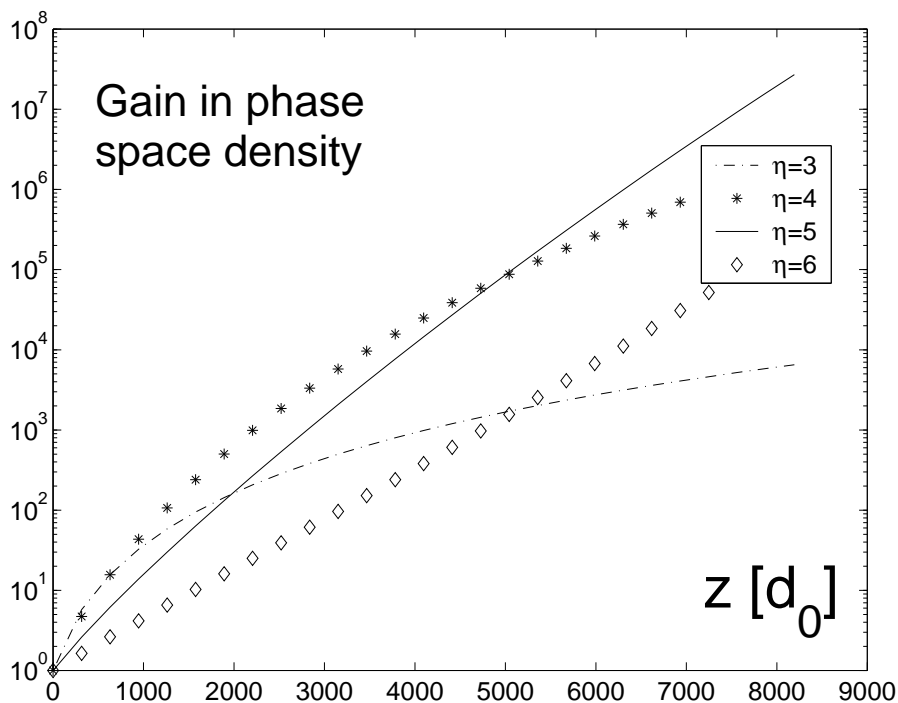}}
\caption{\small For one-dimensional evaporative cooling,
gain in phase space density as a function of the position
$z$ in the magnetic guide. The position is expressed in units
of the initial mean free path  $d_0=\sqrt{\pi}/(2n_0 \sigma)$. These curves are the
numerical solutions
of the hydrodynamic type equations  in the case where $\eta=\epsilon_{\rm evap}(z)/(k_B T(z))$ 
is fixed.}\label{fig:evap}
\end{figure}

We have solved numerically the hydrodynamic type equations in steady state,
assuming that the $z$-dependence of the parameter $\Lambda(z)$ is adjusted to maintain a
$z$-independent ratio $\eta=\epsilon_{\rm evap}(z)/(k_B T(z))$. For the specific
set of parameters of \S\ref{sub:param} we have to gain seven orders
of magnitude on the phase space density to reach quantum degeneracy. The smallest
spatial length of evaporation required is obtained for $\eta\simeq 5$
as shown in figure~\ref{fig:evap}; it is equal
to 7600 $d_0$  where $d_0=\sqrt{\pi}/(2n_0 \sigma)$ is 
the mean free path at the entrance of the magnetic guide,
that is $\simeq 11$ meters for the considered parameters. After 
evaporative cooling along these 11 meters the
flux of particles has been reduced by a factor 90 and the temperature
has been decreased by a factor 4000.

We have also performed a numerical simulation of the full Boltzmann equation
by a Monte Carlo algorithm using macro-atoms \cite{Bird}. The resulting numerical calculations
take several days on a workstation. We have first simulated the one-dimensional 
evaporation scheme. In this way we have confirmed the accuracy of the predictions
based on the ansatz Eq.(\ref{eq:ansatz}). We have also performed simulations
for the full two-dimensional evaporation. We have found typically that the length
required to reach quantum degeneracy is reduced by a factor three as compared to
one-dimensional evaporation, with the same loss of two orders of magnitude
on the flux. For the specific set of parameters of \S\ref{sub:param}
the required evaporation length reaches the experimentally reasonable value
of 4 meters.

\section{Coherence properties in the ideal Bose gas model} \label{ideal}
We now assume that evaporative cooling has allowed to reach the quantum degenerate regime at
a certain distance from the entrance of the tube. We have not performed any kinetic study
of the approach of quantum degeneracy, and we will here simply assume that the gas
is at thermal equilibrium in the frame moving at the mean velocity of the gas. Such an assumption
is reasonable if the temperature remains significantly larger than the quantum $\hbar\omega_\perp$
of transverse oscillation of the atoms of the guide. If the temperature was much smaller than $\hbar\omega_\perp/k_B$
the transverse degrees of freedom of the gas would be frozen in the ground state of the transverse harmonic
oscillator; the gas would become a free one-dimensional Bose gas along $z$, that could not thermalize as colliding
identical particles in one dimension simply exchange their momenta.

In this section we consider the model of the ideal Bose gas. The effects of atomic interactions
are discussed in the next section.

\subsection{Transverse Bose-Einstein condensation}
Let us enclose the ideal Bose gas in a fictitious box of size $L$ along $z$, with
periodic boundary conditions. Transversally the gas is confined by the harmonic potential
Eq.(\ref{eq:pot}). The one-particle eigenstates of the system are then labeled by three integers,
the non-negative integers $l_x,l_y$ labeling the eigenstates of the harmonic oscillator along $x$ and
$y$, and the integer $l_z$ labeling the momentum along $z$:
\begin{equation}
\hbar k_z = \frac{2\pi}{L} l_z.
\label{eq:genial}
\end{equation}
In the grand canonical ensemble the mean occupation number of the 
single particle state $\vec{l}=(l_x,l_y,l_z)$
is given by
\begin{equation}
n(\vec{l},\mu) = \left\{\exp\left[\beta\left(\frac{\hbar^2k_z^2}{2m}+(l_x+l_y)\hbar\omega_\perp-\mu\right)\right]-1\right\}^{-1}.
\label{eq:Bose}
\end{equation}
For convenience we have included the transverse zero-point motion energy in the chemical potential,
so that $\mu$ varies from $-\infty$ to 0.

It turns out that in our trapping geometry there is no Bose-Einstein condensation in the
thermodynamical limit defined as $L,N\rightarrow +\infty$ with a fixed linear density $N/L$, $N$ being the mean
number of particles. 

Let us consider indeed a fixed value of $L$ and let us define the maximal mean
number of particles $N'_{\rm max}$ that can be put in all states but the ground state of the trap
at a fixed temperature. A Bose-Einstein condensate forms in the ground state of the trap
when $N$ exceeds $N'_{\rm max}$. As each $n(\vec{l},\mu)$ (for $\vec{l}\neq\vec{0}$) reaches its
maximal accessible value for $\mu=0$, $N'_{\rm max}$ is given by
\begin{equation}
N'_{\rm max} = \sum_{\vec{l}\neq\vec{0}} n(\vec{l},\mu=0).
\end{equation}
A lower bound on $N'_{\rm max}$ is obtained by restricting the sum to the states transversally in the ground
state of the harmonic oscillator:
\begin{equation}
N'_{\rm max} \geq \sum_{l_z\neq 0} \left[\exp\left(\beta\frac{h^2 l_z^2}{2 m L^2}\right) -1\right]^{-1}.
\end{equation}
When $L$ is large enough so that $k_B T \gg h^2/(2m L^2)$ the exponential in the denominator can be expanded to
first order, leading to
\begin{equation}
N'_{\rm max} \geq \frac{\pi}{3} \left(\frac{L}{\lambda}\right)^2
\end{equation}
where we have introduced the thermal de Broglie wavelength 
\begin{equation}
\lambda = \frac{h}{(2\pi m k_B T)^{1/2}}.
\label{eq:lambda}
\end{equation}
One then realizes that $N'_{\rm max}$ grows faster than $N$ in the thermodynamical limit $L\rightarrow +\infty$,
with $N/L$ fixed. This is connected to the known fact that there is no Bose-Einstein
condensation in a one-dimensional homogeneous Bose gas in the thermodynamical limit.

On the other hand, transverse Bose-Einstein condensation 
\cite{Ketterle_ibg} is taking place in our system.
Let us calculate indeed the maximal number of atoms that can be put 
for a fixed temperature in the transversally excited single particle states:
\begin{equation}
N^{\perp}_{\rm max} = \sum_{(l_x,l_y)\neq (0,0)} \sum_{l_z} n(\vec{l},\mu=0).
\end{equation}
Replacing in the large $L$ limit the sum over $l_z$ by an integral and expanding $1/(\exp(x)-1)$ in powers of $\exp(-x)$
we obtain:
\begin{equation}
N^{\perp}_{\rm max} \simeq \frac{L}{\lambda} \sum_{s\geq 1} s^{-1/2} \left\{
\left[1-\exp\left(-s\beta\hbar\omega_\perp\right)\right]^{-2} -1\right\} \simeq \frac{L}{\lambda} \left(\frac{k_B T}{\hbar\omega_\perp}\right)^2
\zeta(5/2).
\label{eq:inter}
\end{equation}
The last equality is correct in the limit $k_B T \gg \hbar\omega_\perp$. The function $\zeta$ is the Zeta function of
Riemann, and $\zeta(5/2)\simeq 1.341$. If the linear density of the gas $\rho_{\rm lin}=N/L$ exceeds the critical value
\begin{equation}
\rho_{\rm lin}^{(c)} = N^{\perp}_{\rm max}/L \simeq \frac{1}{\lambda} \left(\frac{k_B T}{\hbar\omega_\perp}\right)^2\zeta(5/2)
\label{eq:crit}
\end{equation}
the excess of density will accumulate in the transverse ground state of the trap.
At linear densities much higher than $\rho_{\rm lin}^{(c)}$ the gas becomes almost monomode transversally:
this is an interesting feature for atom optics applications as we have achieved in this way a Heisenberg limited
transverse focalization of the beam.

We can also consider the maximal on-axis density of atoms in the excited transverse states.
In this case it is
more convenient to label the eigenstates of the two-dimensional harmonic oscillator by the angular
momentum $M$ along $z$ and the radial quantum number $l_{\rm r}$. Then the wavefunctions on
$z$ axis, that is in $x=y=0$, have
a squared modulus equal to zero if $M\neq 0$ and equal to $m\omega_{\perp}/(\pi \hbar)$ for $M=0$.
We recall that the states with $M=0$ have an energy above the zero-point energy given by
$2 l_{\rm r} \hbar\omega_\perp$.
We then obtain the maximal on-axis density of atoms in excited transverse states:
\begin{equation}
\rho_{\rm axis}^{(c)} = \frac{m\omega_{\perp}}{\pi \hbar L} \sum_{l_{\rm r}\geq 1} \sum_{l_z}
\left\{\exp\left[\beta\left(\frac{h^2 l_z^2}{2m L^2}+2 l_{\rm r} \hbar\omega_\perp\right)-1\right]\right\}^{-1}.
\label{eq:onaxis1}
\end{equation}
With the same algebraic transformations as for Eq.(\ref{eq:inter}) we obtain
\begin{equation}
\rho_{\rm axis}^{(c)} \simeq \frac{m\omega_{\perp}}{\pi \hbar \lambda} \sum_{s\geq 1} s^{-1/2}
\left[\frac{1}{1-\exp(-2 s\beta\hbar\omega_\perp)}-1\right]
\simeq \frac{\zeta(3/2)}{\lambda ^3}
\label{eq:onaxis2}
\end{equation}
where $\zeta(3/2)\simeq 2.612$ and we have used $k_B T \gg \hbar\omega_\perp$.
Transverse condensation therefore takes place when the usual Einstein's condition $\rho \lambda^3 =
\zeta(3/2)$ is satisfied on the axis of the trap!

\subsection{In the quantum degenerate regime}
We assume now that the gas is in the strongly degenerate regime, with a linear density $\rho_{\rm lin}$
larger than the critical value Eq.(\ref{eq:crit}). The linear density of atoms in the excited transverse states
has reached its saturated value $\rho_{\rm lin}^{(c)}$.
This implies that $|\mu|$ is much smaller than $\hbar\omega_\perp$ ($\mu$ can be replaced by
zero for the transversally excited states); as $k_B T \gg \hbar\omega_\perp$ 
one has also 
\begin{equation}
|\mu|\ll k_B T
\label{eq:crucial}
\end{equation}
so that the occupation number
of the absolute trap ground state $\vec{l}=\vec{0}$, though not of order $N$, is much larger than unity:
\begin{equation}
n(\vec{l}=0) = \frac{1}{\exp(-\beta\mu) -1} \gg 1.
\end{equation}
Let us calculate the linear density of atoms in the transverse ground state of the trap:
\begin{eqnarray}
\rho_{\rm lin}^{(0)} &=& \frac{1}{L} \sum_{l_z} n(0,0,l_z;\mu) \\
&=& \frac{1}{L} \sum_{l_z} \left\{\exp[\beta(\hbar^2 k_z^2/(2m) -\mu)]-1\right\}^{-1}.
\end{eqnarray}
As $\beta |\mu|\ll 1$ and the sum is one-dimensional the main contribution to the sum comes
from states with kinetic energies on the order of $|\mu|$. Expanding the exponential in the Bose formula
to first order we obtain a Lorentzian approximation for the occupation number as function
of momentum:
\begin{equation}
n(0,0,l_z;\mu) \simeq \frac{k_B T}{\hbar^2 k_z^2/(2m) + |\mu|}.
\label{eq:lorentz}
\end{equation}
Replacing finally the sum by an integral we obtain
\begin{equation}
\rho_{\rm lin}^{(0)} = \frac{1}{\lambda} \left(\frac{\pi k_B T}{|\mu|}\right)^{1/2}.
\label{eq:rho_lin_0}
\end{equation}
This allows to express the chemical potential as function of density, when combined with the
relation 
\begin{equation}
\rho_{\rm lin}=\rho_{\rm lin}^{(c)} + \rho_{\rm lin}^{(0)}.
\end{equation}
Note that such a calculation would fail for a two-dimensional or three-dimensional free
Bose gas, a Lorentzian momentum distribution being not normalizable in this case.

To characterize the coherence properties of the gas we use correlation functions  for
the atomic field operator $\hat{\Psi}(x,y,z)$ in direct analogy  with the correlation
functions considered in optics for the photonic field \cite{quantum_optics}.

\begin{figure}[htb]
\epsfxsize=12cm \centerline{\epsfbox{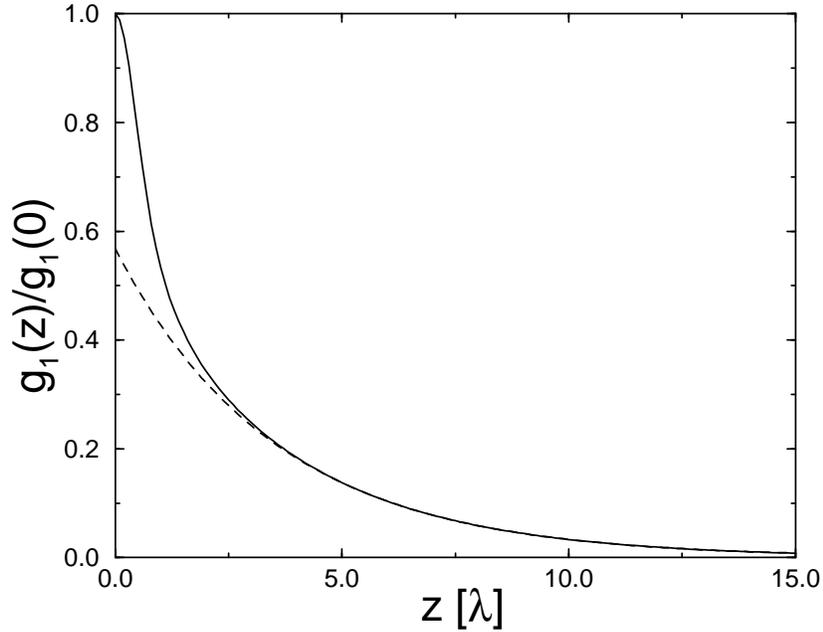}}
\caption{\small Correlation function $g_1(z)$ for the ideal Bose gas in the thermodynamic limit. 
The temperature is $k_B T = 20 \hbar \omega_\perp$ and the linear density is $\rho_{\rm lin}= 1.1 \rho_{\rm lin}^{(c)}$
where the critical density
$\rho_{\rm lin}^{(c)}$ is calculated with the approximate formula Eq.(\ref{eq:crit}). 
One finds numerically $\beta \mu \simeq -6.4\times 10^{-3}$.
Solid line: full
correlation function, calculated numerically. Dashed line: 
contribution of the atoms in the transverse ground state of the trap,
from the analytical formulas Eq.(\ref{eq:rho_lin_0}) 
and Eqs.(\ref{eq:avant},\ref{eq:plus_exacte}). The coherence length 
is found to be $l_c \simeq 3.5 \lambda$.}\label{fig:g1}
\end{figure}

We define the first order correlation function as
\begin{equation}
g_1(z) = \langle \hat{\Psi}^\dagger (0,0,z) \hat{\Psi}(0,0,0)\rangle
\end{equation}
where the expectation value is taken in thermal equilibrium; this function is sensitive
to the coherence of the atomic field between two points on the axis of the guide separated
by a distance $|z|$. It can be written as the sum of the contributions of the transversally excited
states and of the states in the transverse ground state.
The two contributions behave in a very different way in the degenerate limit, that is in the limit
$\mu\rightarrow 0$. This is illustrated in figure~\ref{fig:g1} for a moderately degenerate regime.

By calculations similar to the ones leading to
Eqs.(\ref{eq:onaxis1},\ref{eq:onaxis2}) we find for the contribution of the transverse
excited states to $g_1$ in the thermodynamical limit:
\begin{equation}
g_1^{\perp}(z) = \frac{m\omega_\perp}{\pi\hbar\lambda} \sum_{s\geq 1}
\frac{e^{\beta \mu s}}{s^{1/2}}
\left[\frac{1}{1-\exp(-2s\beta\hbar\omega_\perp)}-1\right]
\exp\left(-\pi \frac{z^2}{s \lambda ^2}\right)\label{eq:exc}.
\end{equation}
Using furthermore the fact that $|\mu|<\hbar\omega_{\perp}\ll k_B T$ 
we set $\mu=0$ and we expand
the exponential function between square brackets to first order in 
$\beta\hbar\omega_{\perp}$:
\begin{equation}
g_1^{\perp}(z) \simeq \frac{1}{\lambda^3} \sum_{s\geq 1} \frac{1}{s^{3/2}} \exp\left(-\pi \frac{z^2}{s \lambda ^2}\right).
\end{equation}
The maximal value of $g_1^{\perp}$ is obtained in $z=0$ and is equal to $\rho_{\rm axis}^{(c)}$. The half-width
of $g_1^{\perp}$ is on the order of $0.75 \lambda$.

The contribution to $g_1$ of the states in the transverse ground state is given by
\begin{equation}
g_1^{(0)}(z) = \frac{m\omega_\perp}{\pi\hbar L} \sum_{l_z} n(0,0,l_z;\mu) e^{i k_z z}.
\end{equation}
In the thermodynamical limit, we replace the sum by an integral. We use the Lorentzian approximation for the occupation
numbers Eq.(\ref{eq:lorentz}) and calculate its Fourier transform. This leads to a correlation function
being an exponential function of $|z|$:
\begin{equation}
g_1^{(0)}(z) \simeq \frac{m\omega_\perp}{\pi \hbar}\rho_{\rm lin}^{(0)} e^{-z/l_c}
\label{eq:avant}
\end{equation}
with a coherence length 
\begin{equation}
l_c = \frac{\hbar}{(2m|\mu|)^{1/2} }
= \frac{\lambda^2}{2\pi}\rho_{\rm lin}^{(0)}.
\label{eq:plus_exacte}
\end{equation}
The exponential decay of $g_1$ found here is to be contrasted with $g_1$ going to a finite value 
in the case of a three dimensional homogeneous Bose-Einstein condensate \cite{revue}.
The coherence length $l_c$ can be larger than the thermal de Broglie wavelength $\lambda$. 
It can be expressed in terms of
the on-axis density using the relation $g_1^{(0)}(z=0)=\rho_{\rm axis}-\rho_{\rm axis}^{(c)}
\equiv\rho_{\rm axis}^{(0)}$:
\begin{equation}
l_c = \frac{\lambda^2 \hbar}{2m\omega_\perp}\rho_{\rm axis}^{(0)}.
\end{equation}
A related issue is the expression connecting the on-axis and the linear density of the
atoms in the transverse Bose-Einstein condensate:
\begin{equation}
\frac{\rho_{\rm axis}^{(0)}}{\rho_{\rm axis}^{(c)}} \simeq \frac{2\zeta(5/2)}{\zeta(3/2)}
\frac{k_B T}{\hbar\omega_{\perp}}\ 
\frac{\rho_{\rm lin}^{(0)}}{\rho_{\rm lin}^{(c)}}.
\label{eq:rapport}
\end{equation}
This expression holds in the regime $k_B T \gg \hbar\omega_{\perp}$
so that a modest value of $\rho_{\rm lin}$ above the critical value $\rho_{\rm lin}^{(c)}$
may actually correspond to a strongly degenerate regime 
$\rho_{\rm axis}\lambda^3\gg 2.612$.

We define the second order correlation function of the atomic field as
\begin{equation}
g_2(z) = \langle \hat{\Psi}^\dagger(0,0,z)\hat{\Psi}^\dagger(0,0,0) \hat{\Psi}(0,0,0)
\hat{\Psi}(0,0,z)\rangle.
\end{equation}
From a field point of view $g_2$ is the correlation function of the intensity of the field;
it is a measure of the intensity fluctuations of the field. 
From a corpuscular point of view $g_2(z)$ is proportional to the pair distribution 
function of
the atoms in the gas, that is the probability density to find a pair of atoms separated
by a distance $|z|$ in the gas.

For the ideal Bose gas in the grand canonical ensemble the use of Wick's theorem
readily allows one to express $g_2$ in terms of $g_1$ by performing all the possible
binary contractions of the field operators:
\begin{equation}
g_2(z) = g_1^2(0) + g_1^2(z).
\label{eq:g2p}
\end{equation}
We find the unpleasant result that the gas is subject to large intensity fluctuations over
a length scale on the order of the coherence length $l_c$. In particular 
\begin{equation}
\frac{g_2(0)}{g_1^2(0)}  = 2.
\end{equation}
This value of two is typical of a bosonic bunching effect that manifests itself in an Hanbury-Brown
and Twiss experiment in optics with thermal sources. It has to be contrasted with $g_2 \simeq g_1^2$
obtained with laser light for photons or with almost pure Bose-Einstein atomic condensates
\cite{meanf}. The output of our magnetic guide cannot be termed an `atom-laser' if
$g_2(z)$ significantly differs from $g_1^2(0)$ over the coherence length of the field.
Fortunately we shall see in the next sections that atomic interactions can improve the situation.

\noindent {\bf Remark:}

A careful reader may argue that it is dangerous to use the grand canonical ensemble for the ideal Bose gas
to calculate fluctuations of the number of particles, and therefore of the field intensity.
This fear is justified when a condensate is formed, a well known problem in three-dimensions: the number
of particles in the condensate has then unphysically large fluctuations.
This problem does not take place here in the thermodynamical limit where no Bose-Einstein condensate is formed.
More precisely one can deduce from $g_2$ 
the standard deviation of the total number  of particles:
\begin{equation}
\frac{\Delta N}{N} = \left(\frac{l_c}{L}\right)^{1/2}.
\end{equation}
The relative fluctuations in the number of particles become small as soon as the length $L$ of the gas becomes
much larger than the coherence length $l_c$.

To illustrate the strong intensity fluctuations of the ideal Bose gas in a dramatic way we introduce the
Sudarshan-Glauber P representation of the many-body density operator \cite{quantum_optics}:
\begin{equation}
\hat{\sigma} = \int {\cal D}\Psi
\,  |\mbox{coh}:\Psi\rangle
\langle\mbox{coh}:\Psi|\,  P(\{\Psi\},\{\Psi^*\}).
\label{eq:defP}
\end{equation}
This expression is a functional integral $\int {\cal D}\Psi$ over the real part
and the imaginary part of the c-number field $\Psi(\vec{r}\,)$.
It involves the coherent or Glauber state of the atomic field associated to
$\Psi$:
\begin{equation}
|\mbox{coh}:\Psi\rangle \equiv \exp\left[-\frac{1}{2} \int d^3\vec{r}\, |\Psi(\vec{r}\,)|^2\right]
\exp\left[\int d^3\vec{r}\, \Psi(\vec{r}\,) \hat{\Psi}^\dagger(\vec{r}\,)\right] |\mbox{vacuum}\rangle.
\label{eq:coh}
\end{equation}
Here the many-body density operator $\hat{\sigma}$ is the grand canonical thermal density operator
\begin{equation}
\hat{\sigma} \propto \exp[-\beta(\hat{H}-\mu \hat{N})]
\end{equation}
where $\hat{H}$ is the Hamiltonian of the gas containing the kinetic energy and the trapping potential
energy, and $\hat{N}$ is the operator total number of particles. The Glauber distribution function $P$ can then be
calculated exactly \cite{quantum_optics}. One expands the field $\Psi$ on the eigenmodes of the trap:
\begin{equation}
\Psi(\vec{r}\,) \equiv \sum_{\vec{l}} \alpha_{\vec{l}}\; \phi_{l_x}(x)\, \phi_{l_y}(y)
\frac{1}{L^{1/2}}\,e^{i k_z z}
\end{equation}
where $\phi_n$, $n=0,1,\ldots$  are the normalized eigenfunctions of the 1D harmonic oscillator
of frequency $\omega_\perp$ and where the plane waves along $z$ have a wavevector given by
Eq.(\ref{eq:genial}). Then the Glauber distribution function is simply a product over all modes
of Gaussian distributions with squared widths
given by the occupation number of the modes:
\begin{equation}
P(\{\Psi\},\{\Psi^*\}) \propto \exp\left[-\sum_{\vec{l}} 
\frac{|\alpha_{\vec{l}}\, |^2}{n(\vec{l}\,)}\right].
\end{equation}

As this distribution $P$ is positive the thermal equilibrium $\hat{\sigma}$ can be viewed exactly as a statistical 
mixture of coherent states. One can then {\it imagine} that a given experimental realization of
the Bose gas is a coherent state characterized by a field $\Psi$. This field $\Psi$ is stochastic as
it varies in an unpredictable way from one experimental realization to the other. 
As the coherent state 
$|\mbox{coh}:\Psi\rangle$ is an eigenstate of $\hat{\Psi}(\vec{r}\,)$ with the eigenvalue $\Psi(\vec{r}\,)$
one can check that the correlation functions
$g_1$ and $g_2$ are equal to the following averages over all possible realizations of the 
c-number field:
\begin{eqnarray}\label{eq:sav1}
g_1 (z) &=& \langle \Psi^*(z) \Psi(0)\rangle_{\rm stoch} \\
g_2(z) &=& \langle |\Psi(z)|^2 |\Psi(0)|^2 \rangle_{\rm stoch}.
\label{eq:sav2}
\end{eqnarray}

\begin{figure}[htb]
\centerline{ \epsfxsize=7cm \epsfbox{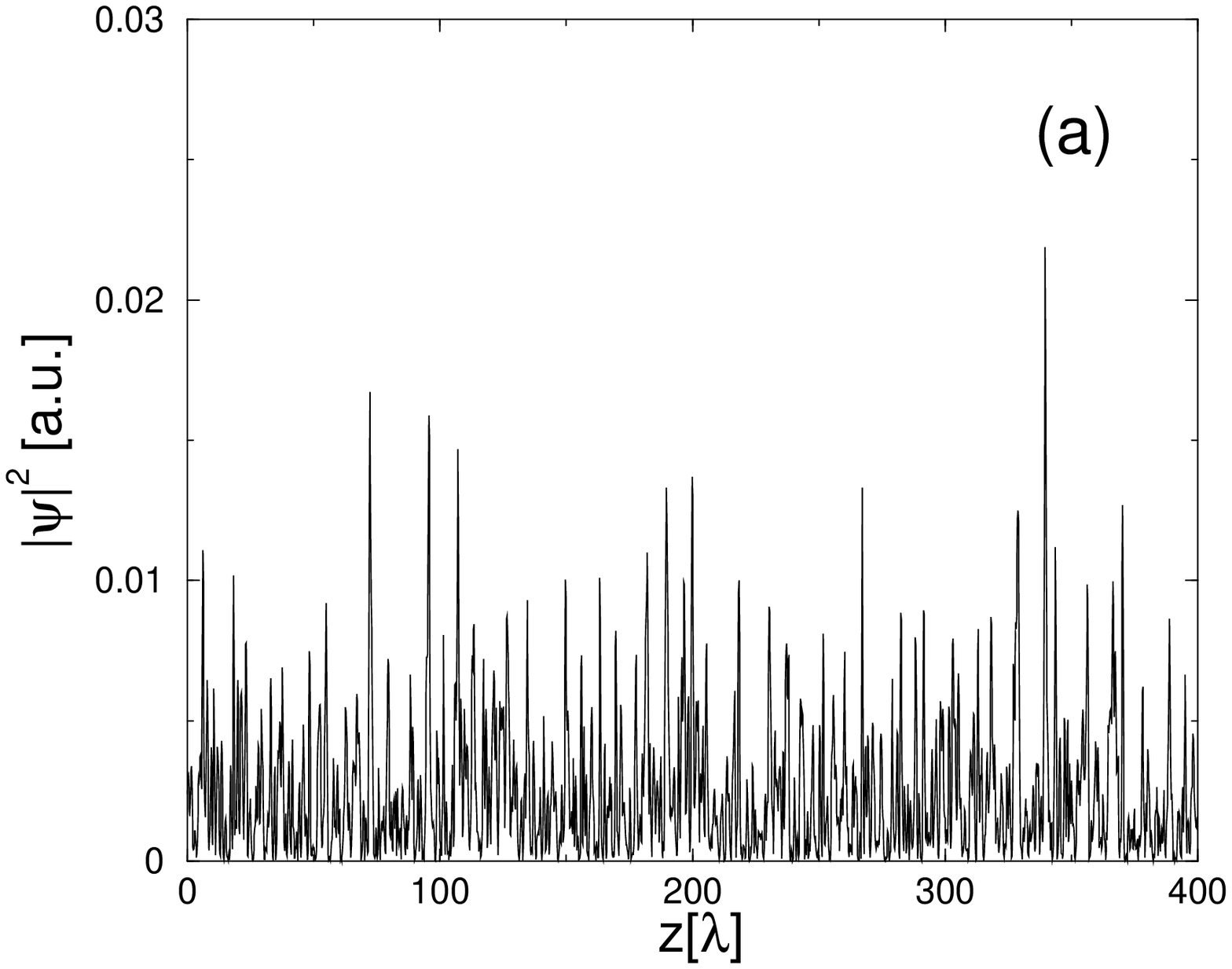}\ \ \ \ \ \ \  \epsfxsize=7cm \epsfbox{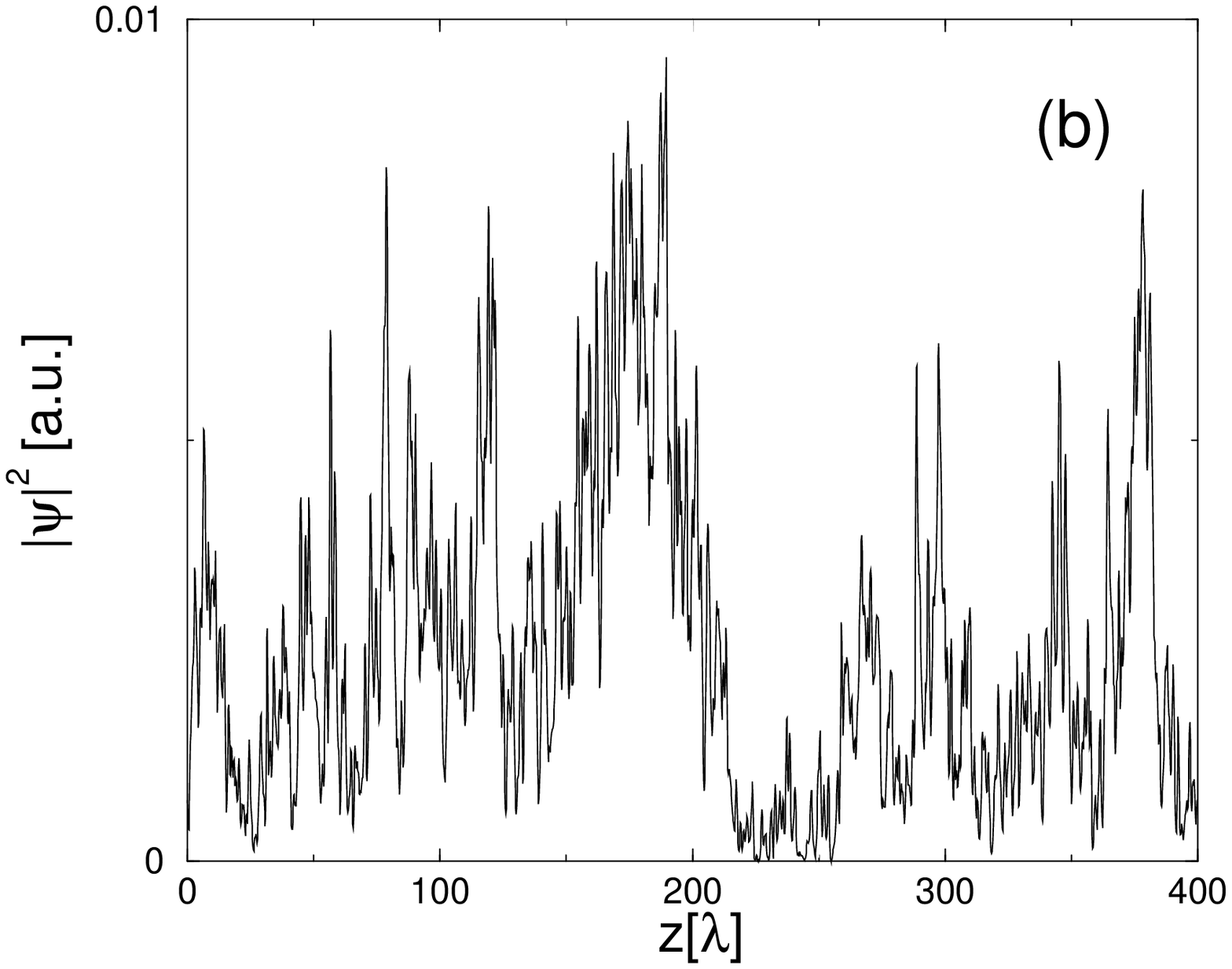}}
\caption{\small Intensity of the atomic field as function of position along $z$ axis for a 
given Monte Carlo sampling  of
the Glauber P distribution at thermal equilibrium. The temperature is $k_B T = 20 \hbar\omega_{\perp}$ and the
size of the box is $L=400\lambda$.
(a) Non transversally condensed regime $\beta\mu = -0.1$ corresponding to $\rho_{\rm lin}=0.92\rho_{\rm lin}^{(c)}$. (b) Strongly degenerate regime 
$\beta\mu = -10^{-4}$ corresponding to $\rho_{\rm lin}=1.4\rho_{\rm lin}^{(c)}$. For (b) the coherence
length is equal to $l_c/\lambda\simeq 28$.
The critical density
$\rho_{\rm lin}^{(c)}$ is calculated with the approximate formula Eq.(\ref{eq:crit}). 
} \label{fig:glaub}
\end{figure}

We have plotted in figure~\ref{fig:glaub} the intensity of the field as function of position for
two numerically generated realizations of $\Psi$. Figure~\ref{fig:glaub}a corresponds to a non-degenerate
situation; the only spatial scale for the intensity fluctuations is the thermal de Broglie wavelength
$\lambda$. Figure~\ref{fig:glaub}b corresponds to a strongly degenerate regime;
there are clearly two spatial scales for the intensity
fluctuations, one on the order of $\lambda$ coming from the transversally non-condensed fraction 
and the other one on the order of $l_c$ due to the gas in the transverse ground state of the trap.  The large
intensity fluctuations at the scale of $l_c$ manifest themselves as ``droplets" 
in the atomic density.

\section{The interacting case: a one-dimensional classical field model} \label{inter}
\subsection{Model Hamiltonian for the transversally Bose condensed gas}
We have seen on the ideal Bose gas model that transverse Bose-Einstein condensation takes place
in the magnetic guide at sufficient high density. Although we have not performed any detailed analysis
we expect the same phenomenon to occur for the interacting Bose gas in the weakly interacting
regime.

At the presently considered low temperatures,
 interaction between the atoms takes place in $s$-wave
mainly, and the relative wavevector of two colliding atoms is much smaller than the inverse of
the scattering length $a_{3d}$ of the interaction potential. The interaction
potential is then commonly replaced by an effective low energy interaction potential
$g_{3d} \delta(\vec{r}_1-\vec{r}_2)$ \cite{Huang,Houches} with a coupling constant
\begin{equation}
g_{3d} =\frac{4\pi\hbar^2}{m} a_{3d}.
\end{equation}
The resulting Hamiltonian is
expressed in terms of the field operator $\hat{\Psi}(x,y,z)$ as
\begin{eqnarray}
\hat{H}_{3d} = \int dx\int dy \int_0^L dz\, \left[
\frac{\hbar^2}{2m}|\vec{\mbox{grad}}\,\hat{\Psi}|^2+\frac{1}{2} g_{3d} \hat{\Psi}^\dagger \hat{\Psi}^\dagger
\hat{\Psi} \hat {\Psi} +(U(x,y)-\mu) \hat{\Psi}^\dagger \hat {\Psi}\right]
\end{eqnarray}
with the trapping potential given in Eq.(\ref{eq:pot}) and with periodic boundary conditions along $z$.
We have included for convenience the term $-\mu \hat{N}$ where $\mu$ is the chemical potential
so that the thermal equilibrium density operator is simply $\propto\exp[-\beta \hat{H}_{3d}]$
in the grand canonical ensemble.

In our magnetic guide geometry the situation is particularly simple 
if the typical interaction energy per particle $\rho_{\rm axis} g_{3d}$ is smaller than the quantum
of transverse oscillation $\hbar\omega_\perp$. 
Transverse Bose-Einstein condensation will then take place
in a transverse wavefunction $\phi_{\perp}(x,y)$ 
close to the ground state of the harmonic oscillator.
If we wish to describe only the one-dimensional Bose gas of atoms in this transverse condensate, assuming that the remaining
atoms have a much smaller density, we can neglect the contribution of the transverse modes to  the field operator by
setting
\begin{equation}
\hat{\Psi}(x,y,z) \simeq \phi_{\perp}(x,y) \hat{\psi}(z).
\label{eq:approx}
\end{equation}
This will eliminate the contribution of the atoms in the transverse excited states of the trap; we keep
in mind however that these atoms are essential to ensure thermalization, with a temperature
$k_B T > \hbar\omega_\perp$.
The reduced field operator $\hat{\psi}$ obeys the usual bosonic commutation relations of a
one-dimensional bosonic field. Inserting the approximate expression for $\hat{\Psi}$ in the
Hamiltonian results
in the following  model Hamiltonian for the transversally condensed Bose gas:
\begin{equation}
\hat{H} = \int_0^L dz\, \left[
\frac{\hbar^2}{2m}|\partial_z\hat{\psi}|^2+\frac{1}{2} g \hat{\psi}^\dagger \hat{\psi}^\dagger
\hat{\psi} \hat {\psi} -\mu \hat{\psi}^\dagger \hat {\psi}\right].
\label{eq:model}
\end{equation}
It corresponds to a one-dimensional Bose gas 
with a contact
interaction potential $g\delta(z_1-z_2)$ between particles with an effective coupling constant 
\cite{Maxim}
\begin{equation}
g = g_{3d} \int\!\!\!\int dx\, dy \, |\phi_\perp(x,y)|^4.
\end{equation}

\subsection{Classical field approximation}

The model Hamiltonian  Eq.(\ref{eq:model}) leads to an exactly solvable $N$-body problem: exact eigenenergies
and eigenfunctions are known \cite{Lieb,Gaudin}. 
It is however not easy to extract information from the exact solution,
even at zero temperature. 

We use here a simpler approach, valid in a sufficiently high temperature regime.
The idea is to write the thermal equilibrium density operator (up to a normalization factor) as the result of
a fictitious time evolution:
\begin{equation}
\frac{d}{d\tau} \hat{\sigma} = -\frac{1}{2} \left(\hat{H}\hat{\sigma} +\hat{\sigma}\hat{H}\right)
\label{eq:master}
\end{equation}
with the initial condition $\hat{\sigma}(\tau=0)$ equal to the identity operator. This evolution corresponds to the so-called imaginary time
evolution. It leads to a density operator at `time' $\tau$ given by:
\begin{equation}
\hat{\sigma} = e^{-\tau \hat{H}}
\end{equation}
which is (up to a normalization factor) the thermal equilibrium density operator at temperature $k_B T =1/\tau$.
High temperatures correspond to low values of $\tau$ that is to short `time' evolution.

To take advantage of the fact that the `time' evolution is short it is better to rewrite Eq.(\ref{eq:master}) using some
of the representations of the density operator introduced in quantum optics. We use here the Glauber P distribution 
already introduced in section \ref{ideal}. It is defined as in Eq.(\ref{eq:defP}) with the difference that
$\psi$ is now a function of the coordinate $z$ only. So we apply the Glauber transform to both
sides of Eq.(\ref{eq:master}). The transform of the right-hand is performed with the following rules:
\begin{eqnarray}
\mbox{Glaub}\,[\hat{\psi}(z)\hat{\nu}] &=& \psi(z) \mbox{Glaub}\,[\hat{\nu}] 
\label{eq:rule1}\\
\mbox{Glaub}\,[\hat{\psi}^\dagger(z)\hat{\nu}] &=& \left(\psi^*(z)-\partial_{\psi(z)}\right) \mbox{Glaub}\,[\hat{\nu}] 
\label{eq:rule2}
\end{eqnarray}
where $\hat{\nu}$ is any operator and $\mbox{Glaub}\,[\hat{\nu}]$ stands for the Glauber P
distribution of $\hat{\nu}$. The first rule Eq.(\ref{eq:rule1}) comes from the fact that the coherent state
$|\mbox{coh}:\psi\rangle$ is an eigenstate of $\hat{\psi}(z)$ with the eigenvalue $\psi(z)$. The second rule Eq.(\ref{eq:rule2})
involves a functional derivative with respect to the field $\psi$, the fields $\psi$ and $\psi^*$ being
formally considered as independent variables. Its derivation closely follows the one
for single mode fields in \cite{quantum_optics}. One first uses the following identity:
\begin{equation}
\partial_{\psi(z)} \left[|\mbox{coh}:\psi\rangle \langle\mbox{coh}:\psi|\right] =
(\hat{\psi}^\dagger(z)-\psi^*(z)) |\mbox{coh}:\psi\rangle \langle\mbox{coh}:\psi| 
\end{equation}
as can be checked from the definition Eq.(\ref{eq:coh})
transposed to the one-dimensional case. 
Then one integrates by part in the functional integral
over the field to convert the derivative over the dyadic $|\mbox{coh}:\psi\rangle \langle\mbox{coh}:\psi|$ into
a derivative of the Glauber P distribution.
We finally obtain the Fokker-Planck type equation for the 
fictitious time evolution of the Glauber distribution:
\begin{equation}
\partial_{\tau} P = - E\, P -\int dz\, \left[\partial_{\psi} (F(z)\, P) 
+\frac{g}{4} \partial_{\psi}^2(\psi^2 P) +\mbox{c.c.}\right].
\label{eq:FP}
\end{equation}

The first term in Eq.(\ref{eq:FP}) involves a multiplication of $P$ by a functional $E$ 
which is simply the Gross-Pitaevskii energy functional \cite{revue}, obtained by replacing in the
Hamiltonian the field operator by the c-number field $\psi$:
\begin{equation}
E[\{\psi\},\{\psi^*\}] = \int_0^L dz\, \left[
\frac{\hbar^2}{2m}|\partial_z{\psi}|^2+\frac{1}{2} g |{\psi}|^4 -\mu |{\psi}|^2\right].
\label{eq:gpen}
\end{equation}
If this term was alone in the evolution equation for $P$ we could readily integrate it to obtain:
\begin{equation}
P_{\rm class}[\{\psi\},\{\psi^*\}] = e^{-\tau E[\{\psi\},\{\psi^*\}]}.
\label{eq:Pclass}
\end{equation}
We have termed this solution $P_{\rm class}$ as it corresponds to the thermal Boltzmann distribution
for a classical field $\psi$ with an energy given by $E[\{\psi\},\{\psi^*\}]$! This is nineteenth century equilibrium 
physics for fields.

The next term in Eq.(\ref{eq:FP}) can be termed a force term by analogy with the Fokker-Planck equation,
as it involves a first order derivative in $\psi$. The `force' functional is given by:
\begin{equation}
F[\{\psi\},\{\psi^*\}](z)=-\frac{1}{2}\partial_{\psi^*(z)} E[\{\psi\},\{\psi^*\}]=
-\frac{1}{2}\left[
-\frac{\hbar^2}{2m}\partial_z^2 + g|\psi|^2 -\mu\right]\psi(z).
\end{equation}
At sufficiently short `time' $\tau$, that is at high temperature,
the effect of the force term during the evolution `time' $\tau$ is to shift the field 
$\psi(z)$ from its initial (random) value by the amount $F(z) \tau$. One may hope that this shift is negligible
for high enough temperature.

Let us calculate this shift for the ideal
Bose gas. The field $\psi$ is then conveniently expanded on plane waves with momenta given by Eq.(\ref{eq:genial}):
\begin{equation}
\psi(z)=\sum_{k_z} \alpha_{k_z} \frac{e^{i k_z z}}{L^{1/2}}.
\end{equation}
Taking now $k_z$ and $\alpha_{k_z}$ as coordinate and field variables (rather than $z$ and $\psi(z)$) we can write:
\begin{equation}
E = \sum_{k_z}\varepsilon_{k_z} \alpha^*_{k_z} \alpha_{k_z} \ \ \ \mbox{and} \ \ \ \ 
F(k_z)=-\frac{1}{2}\partial_{\alpha^*_{k_z}} E = -\frac{1}{2} \varepsilon_{k_z} \alpha_{k_z}
\end{equation}
where we have introduced the mode eigenenergy $\varepsilon_{k_z}=\hbar^2 k_z^2/(2m)-\mu$.
The shift $F(k_z)\tau$ has a negligible effect on field mode $k_z$ if it is small as compared to $\alpha_{k_z}$; this leads to the condition
\begin{equation}
k_B T \gg \varepsilon_{k_z}.
\label{eq:highT}
\end{equation}
The classical field approximation $P_{\rm class}$ is therefore an acceptable approximation for the modes with an energy
much smaller than $k_B T$. The occupation number of such modes can then be obtained from the classical energy equipartition 
formula: a mode of a complex field corresponds to a one-dimensional harmonic oscillator so it has a mean energy equal to
$k_B T$, and
\begin{equation}
\langle |\alpha_{k_z}|^2\rangle = \frac{k_B T}{\varepsilon_{k_z}}.
\end{equation}
This formula coincides indeed with the quantum Bose formula Eq.(\ref{eq:Bose}) in the limit Eq.(\ref{eq:highT}), that is
in the limit of a large occupation number of the mode.

So when can we use the classical field approximation Eq.(\ref{eq:Pclass})? The answer depends on the observable
quantity we wish to calculate. 

For the calculation of the mean energy the classical field approximation
is never acceptable: in the absence of energy cut it predicts an infinite mean energy, the well known
blackbody radiation catastrophe. To save the situation one has to introduce an energy cut $\varepsilon_{\rm cut}$ on the order of 
$k_B T$ to reproduce `by hand' the fact that the Bose formula gives an exponentially small occupation number
to modes with eigenenergy  much larger than $k_B T$. The mean energy is then
finite but depends on the precise value of $\varepsilon_{\rm cut}$.

The conclusion is different for the calculation of the correlation functions $g_1$ and $g_2$.
We have actually already used the classical field approximation in the derivation of $g_1$,
see the approximation Eq.(\ref{eq:lorentz})! This did not lead to any divergence, a fortunate feature
peculiar to the free one-dimensional Bose gas. The classical field predictions for $g_1$ and $g_2$ therefore
do not depend on the energy cut $\varepsilon_{\rm cut}$ provided that the energy cut is large enough. For the
ideal Bose gas the condition is $\varepsilon_{\rm cut}\gg |\mu|$, as the kinetic energy width 
of the Lorentzian Eq.(\ref{eq:lorentz}) is $|\mu|$;
as $\varepsilon_{\rm cut} \sim k_B T$ the classical field calculation of $g_{1,2}$ requires
$k_B T \gg |\mu|$: we recover Eq.(\ref{eq:crucial}).

The validity conditions of the classical field approximation for the interacting case are more
subtle to derive  and will be discussed in \S\ref{subsec:val_cond}.

Finally, the last term in Eq.(\ref{eq:FP}) can be termed a diffusion term by analogy with the Fokker-Planck equation,
as it involves second order derivatives in $\psi$. 
The corresponding `diffusion' matrix in point $z$, given by:
\begin{equation}
D = -\frac{g}{4} \left(\begin{tabular}{cc} $0$ & $\psi^2$ \\ $\psi^{*2}$ & $0$ \end{tabular}\right)
\end{equation}
is however a non positive matrix: 
one can check that the field quadrature along $\psi$ is squeezed by the `diffusion'
while the field quadrature orthogonal to $\psi$ gets anti-squeezed. 
This non-positivity of the `diffusion'
matrix makes it impossible to perform a stochastic, Brownian type simulation of Eq.(\ref{eq:FP}), 
which would have provided
an exact numerical solution to the problem.

\subsection{How to calculate the correlation functions in the classical field approximation}\label{subsec:how}
We take here as Glauber P distribution the classical approximation 
Eq.(\ref{eq:Pclass}), without introducing
 any energy cut, and
we wish to calculate the stochastic averages Eqs.(\ref{eq:sav1},\ref{eq:sav2}). 
This amounts to calculating
ratios of functional integrals over paths parametrized by
$z\in [0,L]$..  For example the first 
order correlation function of the field $\hat{\Psi}(z)$
in the approximation Eq.(\ref{eq:approx}) 
is given by
\begin{equation}
g_1(z) = |\phi_\perp(0,0)|^2\frac{\int d\psi(0)\oint_{\Gamma_{\psi(0)}}{\cal D}\psi\,
\psi^*(z)\psi(0) e^{-\beta E[\{\psi\},\{\psi^*\}]}}
{\int d\psi(0)\oint_{\Gamma_{\psi(0)}} {\cal D}\psi\,
e^{-\beta E[\{\psi\},\{\psi^*\}]}}.
\label{eq:hor}
\end{equation}
In the above formula the functional integrals are performed over all possible
closed paths, as the gas is subject to spatially periodic boundary conditions; we have
split the functional integral as a regular integral over the value of the path
$\psi(0)$ in $z=0$ and a functional integral over the set $\Gamma_{\psi(0)}$ of
all paths starting
with the value $\psi(0)$ in $z=0$ and ending with the same value in $z=L$.
We explain here how to calculate these functional integrals. 
The reader not interested in technicalities may jump directly
to \S\ref{subsec:results}.

To calculate functional integrals like Eq.(\ref{eq:hor})
it is of course possible  
to use a Monte Carlo method to sample the distribution of $\psi$, e.g.\ by representing $\psi$
on a finite spatial grid with step $dz$ and by evolving $\psi$ one time step $dt$ after the other
according to the stochastic evolution 
\begin{equation}
d\psi(z) = -dt\left[-\frac{\hbar^2}{2m}\partial_z^2 + g|\psi|^2 -\mu\right]\psi(z) + \left(\frac{k_B T}{dz}\right)^{1/2} d\xi(z)
\end{equation}
where the $d\xi(z)$'s are statistically independent complex noises of variance $2 dt$. 

There exists 
however a more elegant and much faster solution.  One can use the link between path integrals
and quantum mechanics propagator put forward by Feynman. 
The functional integral over the classical complex field
then corresponds to a propagator in imaginary time of the quantum mechanical problem of a particle
in two dimensions.  The point by point analogy between the two problems is specified in 
the translation table:

\begin{center}
\begin{tabular}{ccc}
{\bf classical field problem}  & & {\bf quantum mechanical analogy} \\
&& \\ 
path integral & $\leftrightarrow $ & quantum propagator \\
abscissa $z$ &$\leftrightarrow $& time $t$ \\
Re\,($\psi(z)$) &$\leftrightarrow $& position $x(t)$ \\
Im\,($\psi(z)$) &$\leftrightarrow $& position $y(t)$ \\
$\int_{\psi(0)=\psi_i}^{\psi(z)=\psi_f}{\cal D}\psi\; e^{-\beta E[\{\psi\},\{\psi^*\}]}$ &$\leftrightarrow $&
$\langle x_f,y_f| e^{-t {\cal H}/\hbar} |x_i,y_i\rangle$ .
\end{tabular}
\end{center}

We have to identify the Hamiltonian ${\cal H}$ of the equivalent quantum mechanics problem.
We postulate the following form:
\begin{equation}
{\cal H} = \frac{p_x^2 + p_y^2}{2M} + V(x,y).
\label{eq:hamile}
\end{equation}
The imaginary time propagator is then expressed in terms of the path integral \cite{Feynman}:
\begin{equation}
\langle x_f,y_f| e^{-t {\cal H}/\hbar} |x_i,y_i\rangle = 
\int_{x(0)=x_i}^{x(t)=x_f} {\cal D}x(\tau) \int_{y(0)=y_i}^{y(t)=y_f} {\cal D}y(\tau)\,
e^{-S[\{x\},\{y\}]/\hbar}
\end{equation}
where the action $S$ is a functional of the path $x(\tau),y(\tau)$:
\begin{equation}
S[\{x\},\{y\}]=\int_0^t d\tau\, \left[\frac{1}{2} M \left(\frac{dx}{d\tau}\right)^2+\frac{1}{2} 
M\left(\frac{dy}{d\tau}\right)^2
+V(x(\tau),y(\tau))\right].
\end{equation}
One identifies this action with $\hbar \beta E$ and one uses the
translation table to obtain the values of the parameters of the equivalent 
quantum mechanics problem:
a mass
\begin{equation}
M = \frac{\hbar^3}{m k_B T}
\end{equation}
and a potential
\begin{equation}
V(x,y) =\hbar\beta \left[ \frac{g}{2} (x^2+y^2)^2-\mu (x^2+y^2)\right].
\label{eq:poten}
\end{equation}
This potential is rotationally invariant, 
as a consequence of the $U(1)$ symmetry of the Hamiltonian
Eq.(\ref{eq:model}). We can therefore classify the eigenstates of $V$ with two quantum numbers, 
an angular momentum $m$ and a radial quantum number $n=0,1,\ldots$. We call $\phi_n^{m}$ the corresponding
normalized eigenvector with eigenvalue $\varepsilon_n^{m}$. As usual the absolute ground state 
of ${\cal H}$ is of angular momentum $m=0$ and radial quantum number $n=0$.

We translate the functional integrals of Eq.(\ref{eq:hor}) into quantum propagators. In particular
we note that the integral
over $\psi(0)$ in Eq.(\ref{eq:hor}) corresponds to an integral over all possible initial  coordinates
of the particle, that is to a trace over all possible initial quantum states of the particle. We finally obtain:
\begin{equation}
g_1(z) =|\phi_\perp(0,0)|^2\frac{\mbox{Tr}\left[ e^{-(L-z){\cal H}/\hbar} (x-iy) e^{-z{\cal H}/\hbar}(x+iy)\right]}
{\mbox{Tr}\left[e^{-L{\cal H}/\hbar}\right]}
\end{equation}
and a similar expression for $g_2$.

Physically, as there is no Bose-Einstein condensation along $z$,
one expects that the length of magnetic guide $L$ in the experiment is much larger than any correlation length of the gas.
One can then take the thermodynamical limit along $z$, putting $L$ to 
infinity while keeping a constant chemical
potential $\mu$ (this ensures that the mean linear density $N/L$ tends to constant)
\cite{math}.
In this case $\exp[-L {\cal H}/\hbar]$
becomes proportional to the projector on the absolute ground state of ${\cal H}$:
\begin{equation}
e^{-L {\cal H}/\hbar} \sim e^{-L \varepsilon_0^{m=0}/\hbar} |\phi_0^{m=0}\rangle
\langle\phi_0^{m=0}|.
\end{equation}
The thermodynamical limit approximation greatly simplifies the expressions for the correlation
functions, as the trace $\mbox{Tr}$ can be restricted to the ground state of ${\cal H}$:
\begin{eqnarray}
\label{eq:g1f}
g_1(z) &=& |\phi_\perp(0,0)|^2 \langle \phi_0^{m=0}| (x-iy) e^{-z({\cal H}-\varepsilon_0^{m=0})/\hbar}(x+iy)|\phi_0^{m=0}\rangle  \\
g_2(z) &=& |\phi_\perp(0,0)|^4\langle \phi_0^{m=0}| (x^2+y^2) e^{-z({\cal H}-\varepsilon_0^{m=0})/\hbar}(x^2+y^2)|\phi_0^{m=0}\rangle.
\label{eq:g2f}
\end{eqnarray}
The operator 
$x+iy$ maps the absolute ground state to a state with angular momentum equal to unity.
If we restrict for simplicity to the large $z$ limit, 
the operator $e^{-z{\cal H}}$ in Eq.(\ref{eq:g1f}) becomes proportional to the projector on the
ground state $\phi_0^{m=1}$ of $\cal H$ with angular momentum $m=1$ so that 
\begin{equation} 
g_1(z)\simeq |\phi_\perp(0,0)|^2 a_1 \exp(-\kappa_1 z)
\end{equation}
with 
\begin{equation}
\kappa_1 = (\varepsilon_0^{m=1}-\varepsilon_0^{m=0})/\hbar \ \ \ \mbox{and}
\ \ \ \ a_1 = |\langle \phi_0^{m=1}|x+i y|\phi_0^{m=0}\rangle|^2.
\label{eq:k1}
\end{equation}
A similar analysis can be applied to the correlation function $g_2$, with the difference that 
the operator $x^2+y^2$ in Eq.(\ref{eq:g2f}) 
maps the absolute ground state to a rotationally invariant state.
In the large $z$ limit we keep the contributions of the first two 
eigenstates with $m=0$ to obtain
\begin{equation} 
g_2(z)\simeq g_1^2(0) + |\phi_{\perp}(0,0)|^4 a_2 \exp(-\kappa_2 z)
\end{equation}
with
\begin{equation}
\label{eq:k2}
\kappa_2 = (\varepsilon_1^{m=0}-\varepsilon_0^{m=0})/\hbar \ \ \ \ \mbox{and}\ \ \ \ 
a_2 = |\langle \phi_1^{m=0}|x^2+ y^2|\phi_0^{m=0}\rangle|^2.
\end{equation}

\subsection{Results of the classical field approximation}\label{subsec:results}
We wish to calculate the correlation functions $g_1$ and $g_2$ using the formalism of \S\ref{subsec:how}.
One has then to solve the quantum mechanics equivalent problem of a particle in two-dimension
with the Hamiltonian Eq.(\ref{eq:hamile}). This Hamiltonian can be diagonalized numerically.
We wish to express the results in terms of the linear density
of the transversally condensed Bose gas, rather than in terms of the chemical potential $\mu$. For a given linear
density $\rho_{\rm lin}^{(0)}$ we therefore have to adjust $\mu$ in order to satisfy
\begin{equation}
\rho_{\rm lin}^{(0)}=\langle \hat{\psi}^\dagger(0)\hat{\psi}(0)\rangle.
\end{equation} 
The problem can be simplified by an efficient parameterization. We 
express the coordinates $x$ and $y$
in the quantum mechanics analogy (which correspond to the real and imaginary part of
$\psi$) in units of the square root of the linear density. We express the physical length $z$ 
in units of $\rho_{\rm lin}^{(0)} \lambda^2/(2\pi)$ where $\lambda$ is the thermal
de Broglie wavelength Eq.(\ref{eq:lambda}). We then find that once $\mu$ has been eliminated there
is a single parameter left in the classical field theory:
\begin{equation}
\chi = \frac{\rho_{\rm lin}^{(0)}g}{k_B T} \, \frac{\left(\rho_{\rm lin}^{(0)}\lambda\right)^2}{2\pi}.
\label{eq:chi}
\end{equation}

\begin{figure}[p]
\centerline{
\epsfysize=6.5cm \epsfbox{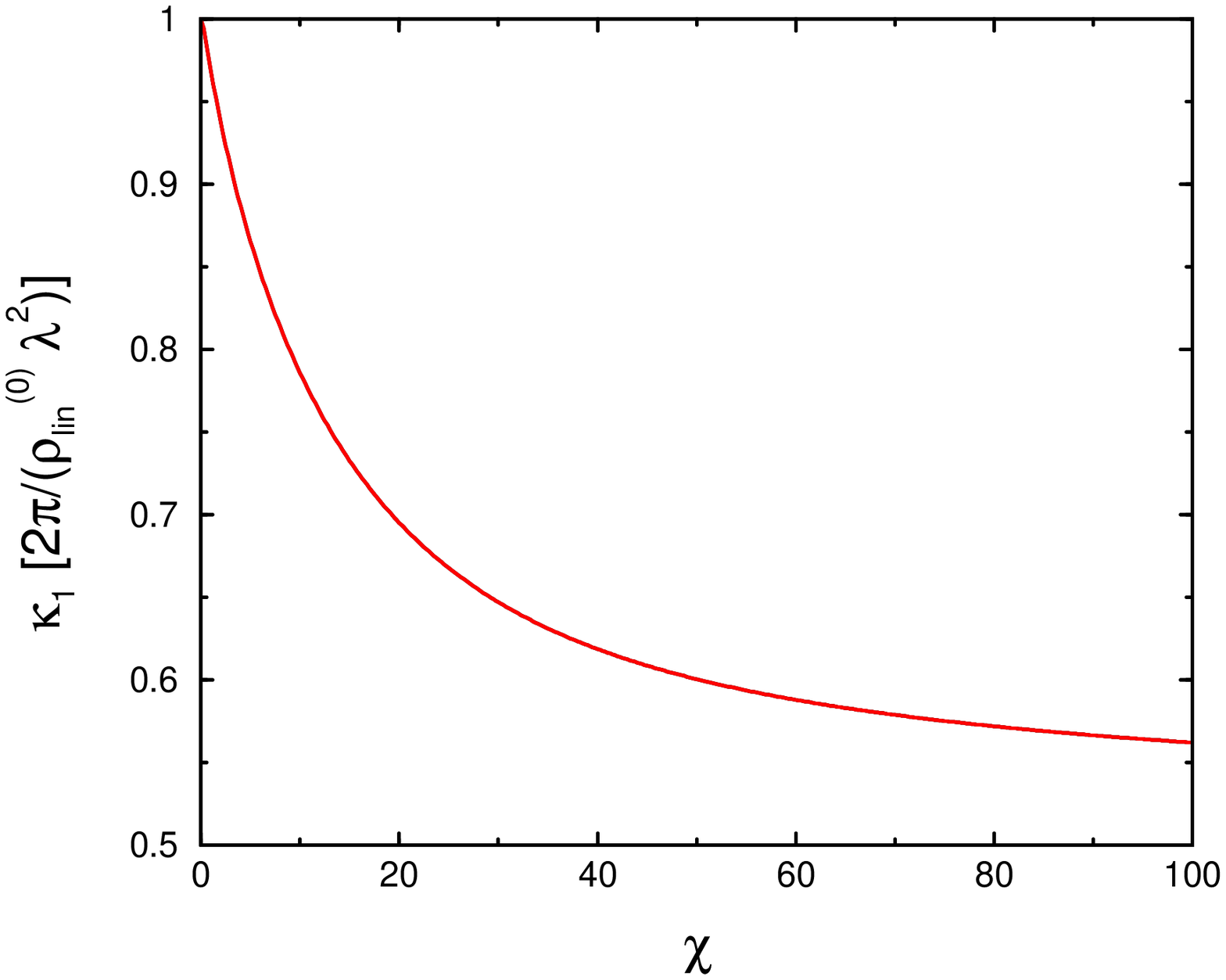}
\ \ \ \ 
\epsfysize=6.5cm \epsfbox{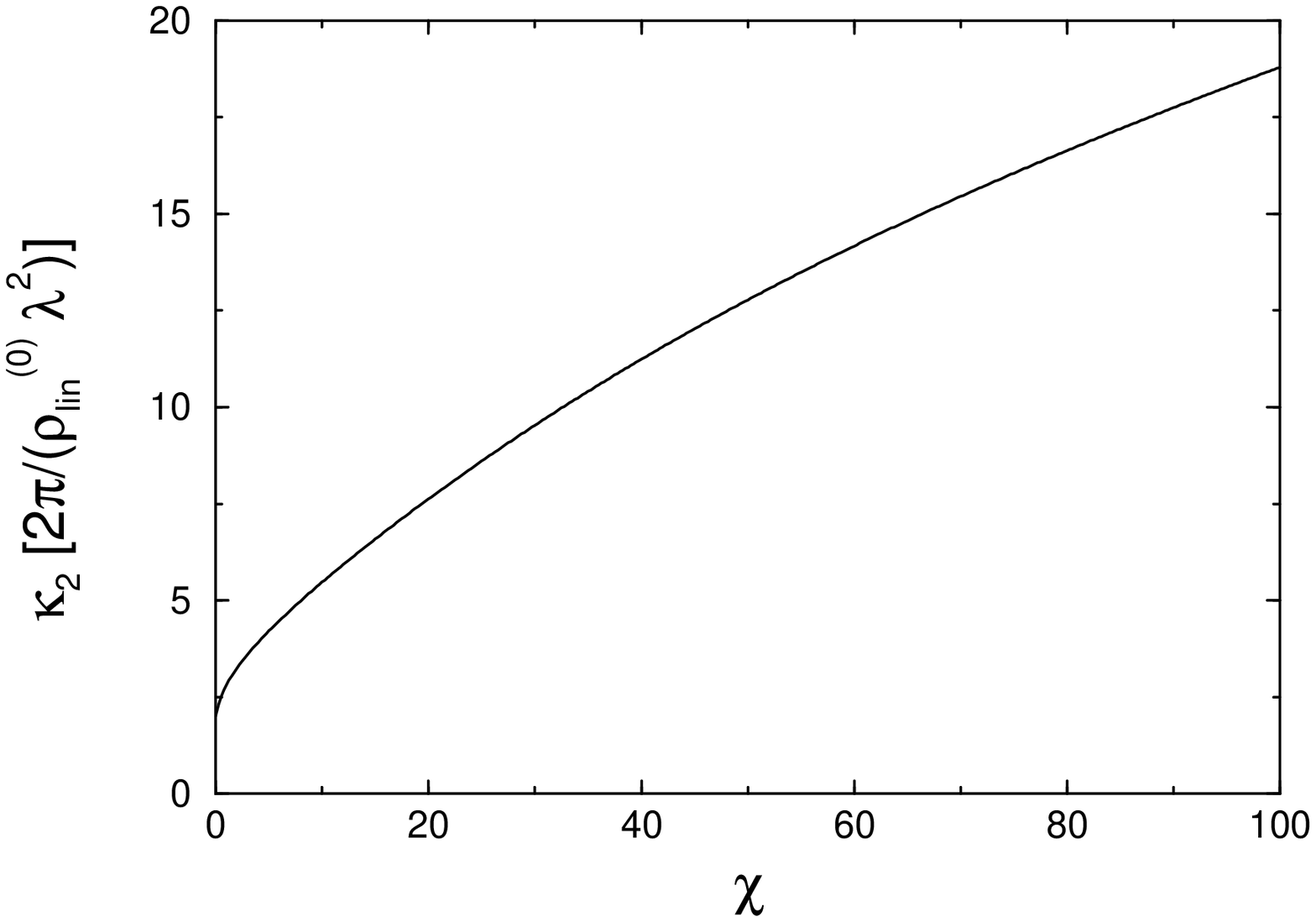}
}
\caption{\small In the one-dimensional classical field model, the inverse field coherence length $\kappa_1$ and 
the inverse intensity correlation length $\kappa_2$ as function of the universal parameter $\chi$ defined in Eq.(\ref{eq:chi}).
 }\label{fig:kappa}
\end{figure}

We plot in figure~\ref{fig:kappa} as function of $\chi$ the coefficients $\kappa_{1,2}$ giving the long distance behavior
of the correlation functions $g_{1,2}$ as defined in Eqs.(\ref{eq:k1},\ref{eq:k2}). In the limit of a vanishing
$\chi$ we recover the results of the ideal Bose gas, Eq.(\ref{eq:plus_exacte}), with $\kappa_1=1/l_c$,
and Eq.(\ref{eq:g2p}) leading to $\kappa_2 = 2 \kappa_1$. For an increasing interaction strength, 
$\chi$ increases:
the coherence length $1/\kappa_1$ has a modest increase by
up to a factor two; 
the correlation length of the intensity fluctuations $1/\kappa_2$ is
dramatically reduced by the atomic interactions and becomes much smaller than the coherence length, a positive
point already~! 

\begin{figure}[p]
\centerline{ \epsfxsize=12 cm \epsfbox{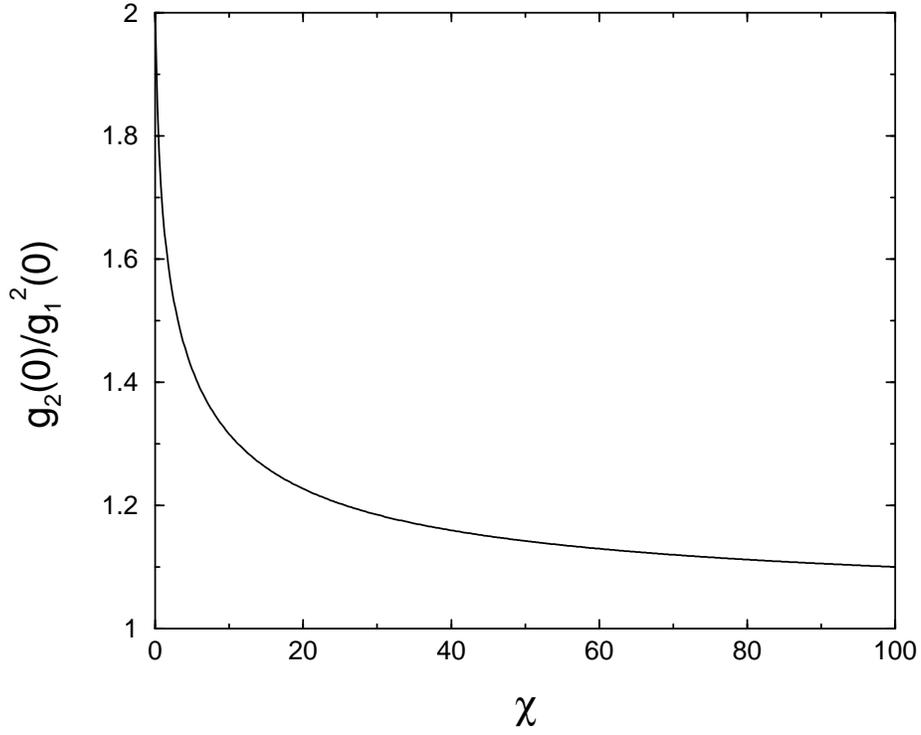} }
\caption{\small In the one-dimensional classical field model, 
indicator of the intensity fluctuations of the
field as function of the universal parameter $\chi$ defined in Eq.(\ref{eq:chi}).}\label{fig:fluc}
\end{figure}

What happens to the amplitude of the intensity fluctuations~? 
In an ideal `atom-laser' there is no fluctuation of the field intensity; the deviation from this ideal situation
can be measured by the ratio of the maximal to the minimal value
of $g_2$, that is 
\begin{equation}
\frac{g_2(0)}{g_1^{2}(0)}=\frac{\langle \phi_0^{(m=0)}| (x^2+y^2)^2|\phi_0^{(m=0)}\rangle}
{\left(\langle \phi_0^{(m=0)}|x^2+y^2|\phi_0^{(m=0)}\rangle\right)^2}.
\label{eq:ratio}
\end{equation}
We have plotted this quantity in figure~\ref{fig:fluc}. It is equal
to two for the ideal Bose gas, as predicted by Eq.(\ref{eq:g2p}). It is sharply reduced by the atomic
interactions for low values of $\chi$ then it slowly goes to unity for large values of $\chi$.

\begin{figure}[htb]
\centerline{
\epsfxsize=7cm \epsfbox{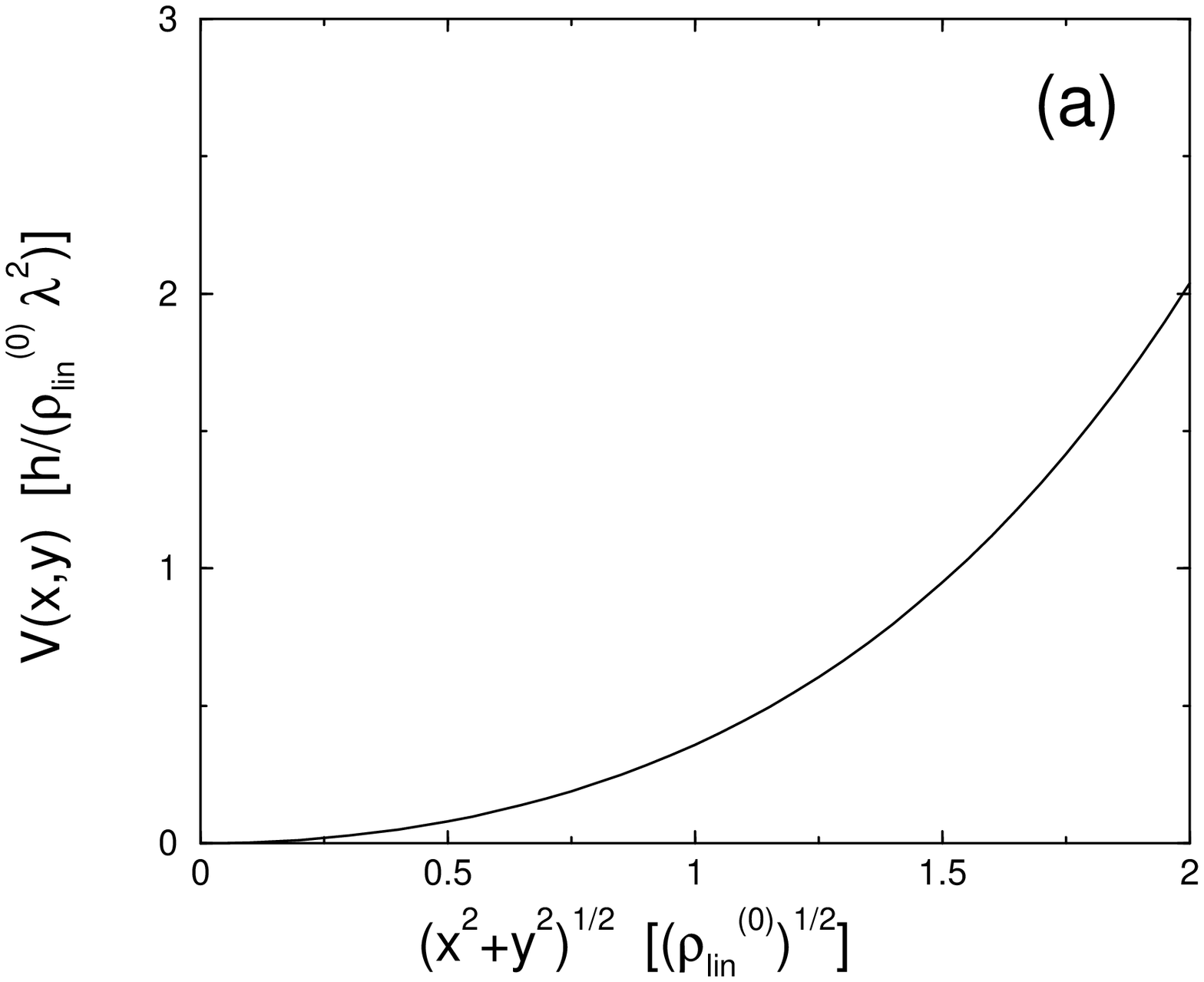}
\ \ \ \ \ \ \ 
\epsfxsize=7cm \epsfbox{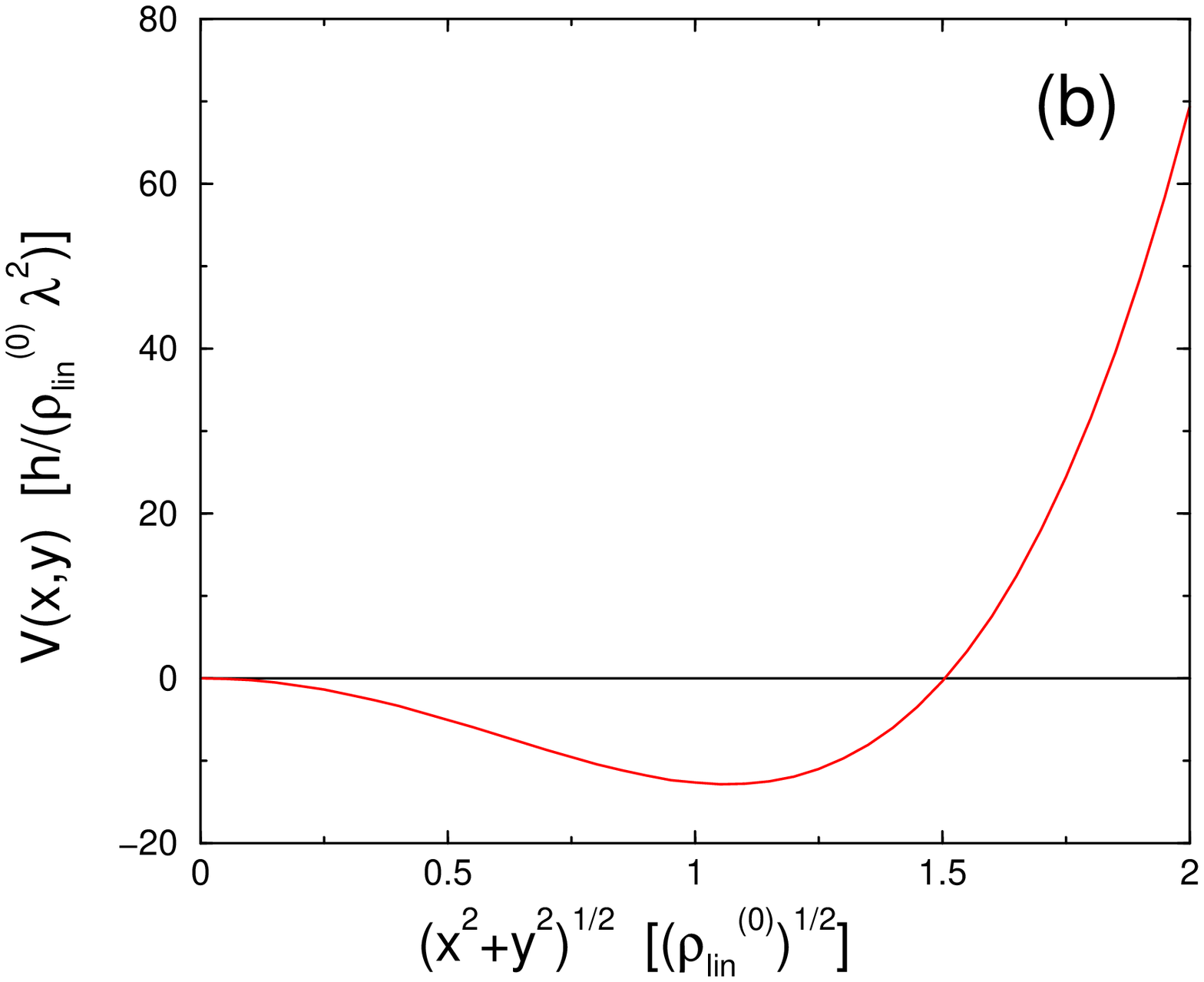}
}
\caption{\small In the quantum mechanics equivalent to the classical field problem, potential $V(x,y)$ seen by the
quantum mechanical particle.
(a) For a low value of $\chi=0.1$ and (b) for a large value of $\chi=20$. 
The units on the axes are such that $V$ depends only on the parameter
$\chi$. }\label{fig:poten}
\end{figure}

Can we understand the origin of the reduction of intensity fluctuations using the quantum mechanics
analogy~? According to Eq.(\ref{eq:ratio}) this amounts to understanding the fluctuations of the distance
of the quantum mechanical particle from the origin in the ground state of ${\cal H}$. By inspection of Eq.(\ref{eq:poten})
giving the trapping potential seen by the quantum mechanical particle we realize that there are two
situations depending on the sign of the chemical potential. 
For a negative chemical potential,
as in the case of the ideal Bose gas, the potential $V$ has an absolute {\it minimum} in $x=y=0$ so that $|\phi_0^{m=0}|^2$
is localized around the origin (see figure~\ref{fig:poten}a): large fluctuations of the intensity of the field
are expected. For a positive chemical potential, which is the case for strong enough repulsive interactions,
the potential $V(x,y)$ has a Mexican hat shape: it has a local {\it maximum} at the origin and a global minimum
on a finite circle (see figure~\ref{fig:poten}b). The ground state wavefunction then tends to be localized around the circle.
The critical regime for the apparition of the Mexican hat potential is such that $\mu=0$; we find numerically
that this corresponds to
\begin{equation}
\chi_c \simeq 0.28.
\end{equation}
This low value of $\chi$ explains why a sharp variation is obtained at the scale of 
figure~\ref{fig:fluc}.

In the large $\chi$ limit the quantum mechanical particle will get more deeply bound at the bottom of the Mexican hat
potential. Writing Schr\"odinger's equation for $\phi_n^{m=0,1}$ in polar coordinates 
and treating perturbatively
the deviation of the Mexican hat plus centrifugal potential from a harmonic approximation we obtain 
after some algebra the large $\chi$'s expansions:

\begin{equation}
\begin{matrix}
\kappa_1 & =& \displaystyle\frac{2\pi}{\rho_{\rm lin}^{(0)}\lambda^2} \left[\frac{1}{2}
+\frac{1}{2\chi^{1/2}}+\ldots\right] 
 &\ \ \ \ \  & \kappa_2 &=&\displaystyle\frac{2\pi}{\rho_{\rm lin}^{(0)}\lambda^2} 
\left[2\chi^{1/2}+\ldots\right]  \\
a_1 &=&\displaystyle \rho_{\rm lin}^{(0)}\left[1+\ldots\right] & & a_2 &=& 
\displaystyle\left(\rho_{\rm lin}^{(0)}\right)^2
\left[\frac{1}{\chi^{1/2}}+\ldots\right]  \\
\displaystyle\frac{g_2(0)}{g_1^2(0)} &=& \displaystyle1+ \frac{1}{\chi^{1/2}}+\ldots & & 
\mu&=&\displaystyle\rho_{\rm lin}^{(0)} g\left[1+\frac{1}{2\chi^{1/2}}+\ldots\right] 
\end{matrix}
\label{eq:large}
\end{equation}
We note that the correlation length of the intensity fluctuations $1/\kappa_2$  
becomes proportional in the large $\chi$ limit
to the so-called healing 
length $\xi$ of the gas, a crucial parameter in the theory of 
Bose-Einstein condensates \cite{revue}:
\begin{equation}
\kappa_2^{-1} \simeq \frac{1}{2}\left(\frac
{\hbar^2} {m\rho_{\rm lin}^{(0)}g} \right)^{1/2}=
\frac{\xi}{\sqrt{2}}.
\label{eq:healing}
\end{equation}


\subsection{Validity conditions of the classical field approximation} 
\label{subsec:val_cond}
The validity condition of the classical field approximation in general depends on
the observable to be calculated. Here the relevant quantities are the first and second
order correlation functions $g_{1,2}$. 

Let us recall briefly what happens
in the ideal Bose gas case. The correlations functions are then
characterized by a single length, the coherence length $l_c=1/\kappa_1=2/\kappa_2$.
The plane wave modes of the field contributing to $g_{1,2}$ have therefore 
wave vectors on the order of $\kappa_1$ or less; for these modes the validity condition
Eq.(\ref{eq:highT}) reads $k_B T \gg \varepsilon_{\kappa_1}$. The mode eigenenergy
$\varepsilon_{\kappa_1}$ is equal to $\hbar^2 \kappa_1^2/(2m)-\mu$; using 
the value Eq.(\ref{eq:plus_exacte}) of the coherence length $l_c=1/\kappa_1$ we arrive
at \begin{equation}
k_B T \gg |\mu|.
\end{equation}
This result was already obtained in Eq.(\ref{eq:crucial}). It can be rewritten with
Eq.(\ref{eq:rho_lin_0}) as a hierarchy among the three relevant lengths of the problem:
\begin{equation}
\left(\rho_{\rm lin}^{(0)}\right)^{-1} < \lambda < l_c=\kappa_1^{-1}.
\end{equation}

What happens in the interacting regime $\chi > 1$~? A difficulty is that the Hamiltonian for
the classical field $\psi$ in Eq.(\ref{eq:gpen})
is no longer quadratic in $\psi$ so that it is not straightforward
to calculate energy eigenmodes. 
Can we find some good quadratic approximation to it?

One could think to use the Bogolubov approach \cite{revue}. In this
approach one identifies the field $\psi_0$ 
minimizing the energy Eq.(\ref{eq:gpen}); in the present 
homogeneous case $\psi_0$ is $z$-independent, $\psi_0=(\mu/g)^{1/2}\exp(i\theta_0)$
where $\theta_0$ is an arbitrary constant phase. Then one
splits the field as $\psi(z)=\psi_0+\delta\psi(z)$. Under the assumption of 
$|\delta\psi|\ll |\psi_0|$ one neglects in Eq.(\ref{eq:gpen}) the
terms cubic and quartic in $\delta\psi$
which leads to a quadratic Hamiltonian that can be diagonalized. This approach is
not well suited to the present situation in the large $L$ limit; it predicts
\begin{equation}
\frac{\langle \delta\psi^*(z)\delta\psi(z)\rangle}{|\psi_0|^2} \simeq
\frac{1}{6} \kappa_1 L.
\end{equation}
Although the emergence of a coefficient proportional
to $\kappa_1$ is promising, the Bogolubov quadratization procedure is not
justified when $L$ exceeds the coherence length of the gas. This is physically
not surprising: the finite range of first order field coherence in our
one-dimensional geometry is precisely due to large fluctuations of
$\psi(z)$ away from $\psi_0$.

Fortunately it is possible to adapt Bogolubov's idea 
taking advantage of the weak intensity
fluctuations of the field in the large $\chi$ regime.
We split the field in a modulus and a phase factor
\begin{equation}
\psi(z) = \rho^{1/2}(z)e^{i\theta(z)}.
\end{equation}
We recall that the intensity $\rho(z)$ and the phase $\theta(z)$ 
of a  field are Hamiltonian
conjugate variables in the same way as $\psi(z)$ and $\psi^*(z)$ are.
The intensity $\rho(z)$ has only small fluctuations away from the most probable
value  $\rho_0=|\psi_0|^2=\mu/g$:
\begin{equation}
\rho(z) = \rho_0 + \delta\rho(z) \ \ \ \ \mbox{with}\ \ \ \ \ 
|\delta\rho(z)|\ll \rho_0.
\end{equation}
The mean value of $\rho(z)$ is simply the linear density of atoms
in the transverse ground state $\rho_{\rm lin}^{(0)}$; to lowest order,
$\rho_0$ can be identified with this mean density.
The phase $\theta$ has on the contrary large fluctuations away from any fixed
constant phase $\theta_0$ over distances larger than a few coherence lengths. 
So contrarily to Bogolubov's method we do not assume $\theta(z)-\theta_0$ to
be small.

The potential energy density of the field $g|\psi|^4/2 - \mu |\psi|^2$ leads to
a constant term and a term quadratic in $\delta\rho$.
In the kinetic energy density of the field we replace $\rho(z)$ by its lowest
order approximation giving a non-zero contribution.
Up to a constant, this leads to a quadratic approximation for the 
Hamiltonian  Eq.(\ref{eq:gpen}):
\begin{equation}
E_{\rm quad} = \int_0^L dz\, \left\{\frac{\hbar^2}{2m}\left[\rho_0
\left(\partial_z\theta\right)^2+\frac{1}{4\rho_0}
\left(\partial_z \delta\rho\right)^2
\right]+\frac{g}{2}\delta\rho^2(z)\right\}.
\end{equation}
One has finally to diagonalize this quadratic Hamiltonian using canonically
conjugate variables.
An easy way is to expand the field variables on the eigenmodes
of the linear time evolution equations:
\begin{eqnarray}
\hbar\partial_t \delta\rho(z) &=& \partial_{\theta(z)} E_{\rm quad} = -\frac{\hbar^2}{m}\rho_0
\partial_z^2\theta(z) \\
-\hbar\partial_t \theta(z) &=& \partial_{\delta\rho(z)} E_{\rm quad} = -\frac{\hbar^2}{4m\rho_0}
\partial_z^2\delta\rho(z)+g\delta\rho(z).
\end{eqnarray}
These well known equations have the form of linearized hydrodynamics equations 
for a superfluid \cite{PRL_stringari}.
Eigenmodes are plane waves of wavevector $k_z$ obeying Eq.(\ref{eq:genial})
and with eigenenergy
\begin{equation}
\varepsilon_{k_z}= \left[\frac{\hbar^2 k_z^2}{2m}\left(\frac{\hbar^2 k_z^2}{2m}
+2\rho_0 g\right)\right]^{1/2}.
\end{equation}
This is the famous Bogolubov spectrum;
the field energy will then formally appear as a sum of decoupled
harmonic oscillators, corresponding to an ideal Bose gas of quasi-particles
\cite{revue}.

We can finally reproduce the reasoning performed above in the ideal Bose gas case
to identify the validity condition of the classical field approximation.
The temperature must be much larger than
the mode eigenenergies $\varepsilon_{k_z}$ at wavevectors $k_z=\kappa_1,\kappa_2$
relevant for the correlation functions $g_1,g_2$:
\begin{equation}
k_B T \gg \varepsilon_{\kappa_1},\varepsilon_{\kappa_2}.
\end{equation}
As $\kappa_2\gg\kappa_1$ in the large $\chi$ regime, the condition involving
$\kappa_2$ is the most stringent one. From Eq.(\ref{eq:healing}) we find
formally the same condition as the ideal Bose gas case
\begin{equation}
k_B T \gg \mu
\label{eq:voila}
\end{equation}
with now a different expression for the chemical potential,
$\mu \simeq \rho_{\rm lin}^{(0)} g$. We note that in the large $\chi$ regime 
and the high temperature regime Eq.(\ref{eq:voila}) we have the following hierarchy
among the various relevant scales of the problem:
\begin{equation}
\left(\rho_{\rm lin}^{(0)}\right)^{-1} < \lambda < 
\kappa_2^{-1}\simeq \xi  < l_c=\kappa_1^{-1} < a
\end{equation}
where the `one-dimensional scattering length' $a$ is defined as
\begin{equation}
g=\frac{\hbar^2}{ma} .
\end{equation}
The property that the microscopic scale $a$ 
is {\it larger} than the other physical lengths of the problem (with the exception of $L$
of course!) ensures that the one-dimensional Bose gas is a weakly interacting
Bose gas.  The fact that this is a necessary condition 
to a classical field approximation (like the Gross-Pitaevskii equation)
is known for three-dimensional
Bose gases, with the difference that the three-dimensional scattering
length $a_{3d}$ has then to be {\it smaller} than the macroscopic scales of 
the gas \cite{revue}. In particular the small gaseous parameter
is $a_{3d}/\xi$ for three-dimensional Bose gases 
whereas it is $\xi/a$ for one-dimensional Bose gases \cite{Lieb}.
This is not surprising if one realizes that the weak interaction limit $g\rightarrow 0$ leads
to $a\rightarrow \infty$ in one-dimension.

\subsection{Comparison with results in the literature}
The model of the one-dimensional interacting Bose gas has already been studied by
several authors without performing the classical field approximation. 

A first line of thought deals with the exact solvable model of a contact
interaction potential \cite{Lieb,Gaudin}. 
To our knowledge a full calculation of $g_1(z)$ and
$g_2(z)$ at finite temperature has not been performed. 
The value of $g_2(0)$ 
can be found in \cite{Lieb} at zero temperature. Exact results have been
obtained in the strongly interacting regime $g\rightarrow +\infty$:
the correlation function
$g_1(z)$ has been calculated 
in \cite{Tonk} at zero temperature; the correlation functions $g_1(z)$ and $g_2(z)$
at finite temperature have been obtained in \cite{Efetov}.
These regimes are different from the finite
temperature, weakly interacting Bose gas considered in the present paper.
The extension of the calculation of $g_2(0)$ to finite temperature seams feasible
but we do not know any reference.

A second line of thought
is to consider the regime of low intensity fluctuations of the field (our large $\chi$ regime):
one may take advantage of the weakness of the  intensity fluctuations by
a linearization of the equations of motion in the hydrodynamic point of view 
\cite{Kane,Haldane,Monien}, 
a Bogolubov type approach \cite{Schwartz}, a quadratization of action in a path integral
formulation \cite{Efetov,Popov}. 

We have checked that the classical field predictions in the asymptotic limit
$\chi\gg 1$ reproduce the results of e.g.\ \cite{Schwartz} when $(k_B T)^{1/2}\gg 
(\rho_{\rm lin}^{(0)}g)^{1/2}$.
The advantage of the classical field approximation is that it is not
restricted to the large $\chi$ regime so that the transition from the ideal Bose
gas to the strongly interacting case can be studied.

In preparing this article we have discovered that a one-dimensional classical field 
model very similar to our model has been studied in \cite{escalope}, with a
different physical motivation.

\section{Conclusion and perspectives} \label{concl}
In this paper we have discussed several aspects of 
our proposal for the production of a continuous `atom-laser' source,
consisting in evaporatively cooling an atomic beam in a long magnetic guide.
From the classical Boltzmann equation and for
expected typical parameters we have estimated the length required to reach the quantum 
degenerate regime, a few meters for two-dimensional evaporation; the corresponding
loss on the atomic flux is only two orders of magnitude.

We have also characterized the coherence properties of
the output beam once quantum degeneracy is reached. The gas is expected
to experience transverse Bose-Einstein condensation in the guide, 
leading to a transversally monomode output beam.  We have therefore introduced
a one-dimensional model for the gas, that we have solved in a classical
field approximation.
The coherence length 
of the field along the axis of the magnetic guide can be much larger than the
thermal de Broglie wavelength, typically by one order of magnitude.
The intensity fluctuations of the beam are very large for an ideal
Bose gas but are strongly reduced by repulsive atomic interactions
when the healing length $\xi$ of the gas becomes smaller than the coherence
length of the gas. 

Possible extensions of this work are the discussion of superfluidity properties
of the `atom-laser' and a complete three-dimensional modeling of the
interacting gas in the magnetic guide.

We acknowledge useful discussions with Gora Shlyapnikov, Philippe Grangier,
Gordon Baym and Tony Leggett.
We thank Alice Sinatra
for helpful comments on the manuscript. Laboratoire Kastler Brossel is
a unit\'e de recherche de l'\'Ecole normale sup\'erieure et de l'Universit\'e
Pierre et Marie Curie, associ\'ee au CNRS.


\begin{thebibliography}{99}
\bibitem{math}
As the length $L$ appears in the quantum
mechanical formulation only through $\exp[-L {\cal H}/\hbar]$ the previous intuitive assumption 
can be cast in the mathematical condition:
$L (\varepsilon_{\rm exc } -\varepsilon_0^{m=0})/\hbar \gg 1$
where $\varepsilon_{\rm exc }$ is the energy of the 
first excited state of ${\cal H}$. As we shall see this results in $\kappa_1 L\gg 1$
where $1/\kappa_1$ is the coherence length
of the gas.
\end{thebibliography}
\end{document}